%% file: main.tex
\documentclass[11pt]{article}

\usepackage{amsmath, amsfonts, amssymb, amsthm}
\usepackage{color}
\usepackage{url}
\usepackage{qcircuit}
\usepackage{algorithm}
\usepackage{algpseudocode}
\usepackage[colorlinks=true,
            linkcolor=blue,
            urlcolor=blue,
            citecolor=blue]{hyperref}
\usepackage{graphicx}
\usepackage{multirow}
\usepackage{multicol}
\usepackage{tikz}
\usepackage{authblk}

\usepackage{subcaption}

\usepackage{pgfplots}
\pgfplotsset{compat=1.18}
\usetikzlibrary{calc,arrows.meta}

\input{q_macro}

\newtheorem{theorem}{Theorem}[section]
\newtheorem{lemma}[theorem]{Lemma}

\title{Space-Efficient Quantum Algorithm for Elliptic Curve Discrete Logarithms with Resource Estimation}

\author[1]{Han Luo\thanks{Equal contribution.}}
\author[2,3]{Ziyi Yang\protect\footnotemark[1]}
\author[2,3]{Ziruo Wang}
\author[2,3]{Yuexin Su}
\author[2,3]{Tongyang Li\thanks{Corresponding author. Email: tongyangli@pku.edu.cn}}

\affil[1]{Interdisciplinary Information Sciences, Tsinghua University}
\affil[2]{Center on Frontiers of Computing Studies, Peking University}
\affil[3]{School of Computer Science, Peking University}
\date{}

\begin{document}

\maketitle

\begin{abstract}
Solving the Elliptic Curve Discrete Logarithm Problem (ECDLP) is critical for evaluating the quantum security of widely deployed elliptic-curve cryptosystems. Consequently, minimizing the number of logical qubits required to execute this algorithm is a key object. In implementations of Shor's algorithm, the space complexity is largely dictated by the modular inversion operation during point addition.
Starting from the extended Euclidean algorithm (EEA), we refine the register-sharing method of Proos and Zalka and propose a space-efficient reversible modular inversion algorithm. We use length registers together with location-controlled arithmetic to store the intermediate variables in a compact form throughout the computation. We then optimize the stepwise update rules and give concrete circuit constructions for the resulting controlled arithmetic components.
This leads to a modular inversion circuit that uses $3n + 4\lfloor \log_2 n \rfloor + O(1)$ logical qubits and $204n^2\log_2 n + O(n^2)$ Toffoli gates. By inserting this modular inversion component into the controlled affine point-addition circuit, we obtain a space-efficient algorithm for the ECDLP with $5n + 4\lfloor \log_2 n \rfloor + O(1)$ qubits and $O(n^3)$ Toffoli gates. In particular, for a 256-bit prime-field curve, our estimate reduces the logical-qubit count to 1333, compared with 2124 in the previous low-width implementation of H\"aner et al.
\end{abstract}

\input{sec_intro/sec_intro}
\input{sec_prelim/sec_prelim}
\input{sec_EEA/sec_EEA}
\input{sec_circuit/sec_circuit}

\input{sec_resource/sec_resource}

\section*{Acknowledgments}
ZY, ZW, YS and TL are supported by the National Natural Science Foundation of China (Grant Numbers 62372006 and 92365117).

\newcommand{\etalchar}[1]{$^{#1}$}
\newcommand{\arxiv}[1]{arXiv:\href{https://arxiv.org/abs/#1}{\ttfamily{#1}}\?}\newcommand{\arXiv}[1]{arXiv:\href{https://arxiv.org/abs/#1}{\ttfamily{#1}}\?}\def\?#1{\if.#1{}\else#1\fi}

\appendix
\input{app_proof/app_proof}
\input{app_anal/app_anal}

\end{document}

%% file: q_macro.tex
\newcommand{\F}{\mathbb{F}}

\newcommand{\floor}[1]{\left\lfloor #1 \right\rfloor}
\newcommand{\ceil}[1]{\left\lceil #1 \right\rceil}

\newcommand{\ket}[1]{|#1 \rangle}

\newcommand{\QFT}{\mathsf{QFT}}

\newcommand{\triang}{\quad\,\blacktriangleright\ \,}

\newcommand{\gateout}[1]{*+<.6em>{#1\ } \POS ="i",[0,0]+R *{\triang},"i"+UR;"i"+UL **\dir{-};"i"+DL **\dir{-};"i"+DR **\dir{-};"i"+UR **\dir{-},"i" \qw}
\newcommand{\multigateout}[2]{*+<1em,.9em>{\hphantom{#2}} \POS [0,0]="i",[0,0].[#1,0]="e", !C *{#2}, [0,0]+R *{\triang}, "e"+UR;"e"+UL **\dir{-};"e"+DL **\dir{-};"e"+DR **\dir{-};"e"+UR **\dir{-},"i" \qw}

\newcommand{\multigateouttwo}[2]{*+<1em,.9em>{\hphantom{#2}} \POS [0,0]="i",[0,0].[#1,0]="e", !C *{#2}, [0,0]+R *{\triang}, [#1,0]+R *{\triang}, "e"+UR;"e"+UL **\dir{-};"e"+DL **\dir{-};"e"+DR **\dir{-};"e"+UR **\dir{-},"i" \qw}
\newcommand{\multigateouttwolow}[2]{*+<1em,.9em>{\hphantom{#2}} \POS [0,0]="i",[0,0].[#1,0]="e", !C *{#2}, [1,0]+R *{\triang},  [#1,0]+R *{\triang},"e"+UR;"e"+UL **\dir{-};"e"+DL **\dir{-};"e"+DR **\dir{-};"e"+UR **\dir{-},"i" \qw}
\newcommand{\multigateouttwohi}[2]{*+<1em,.9em>{\hphantom{#2}} \POS [0,0]="i",[0,0].[#1,0]="e", !C *{#2}, [0,0]+R *{\triang},  [1,0]+R *{\triang},"e"+UR;"e"+UL **\dir{-};"e"+DL **\dir{-};"e"+DR **\dir{-};"e"+UR **\dir{-},"i" \qw}

%% file: sec_intro/sec_intro.tex
\section{Introduction}

Elliptic curve cryptography (ECC), first proposed in the mid-1980s as an alternative framework for public-key cryptography \cite{koblitz1987elliptic, miller1985use}, has become one of the foundational primitives in modern cryptographic systems.
The security of ECC-based cryptosystems relies on the computational difficulty of the Elliptic Curve Discrete Logarithm Problem (ECDLP).
Specifically, given an elliptic curve defined over a finite field, a base point $P$ on that curve, and another point $Q$ resulting from scalar multiplication of $P$ by an unknown integer $m$, the ECDLP asks for the recovery of the scalar $m$ from the two points.
Despite decades of research, no classical algorithm is known that solves the ECDLP in polynomial time of bit length; the best generic attacks, such as Pollard’s rho method, require $\Theta(\sqrt{p})$ group operations, where $p$ denotes the field size of the curve which is exponentially large \cite{galbraith2016recent}.

Compared to classical public-key systems such as RSA \cite{rivest1978method} and finite-field Diffie-Hellman \cite{diffie1976new}, ECC offers equivalent security with substantially smaller key sizes.
For example, according to NIST recommendations \cite{barker2020nist}, a 256-bit elliptic curve provides a security level of 128 bits, which is comparable to that of RSA with a 3072-bit modulus.
This efficiency advantage has led to the widespread adoption of ECC in both theoretical cryptographic constructions and real-world applications. Specifically, ECC is used for key exchange \cite{diffie1976new} and digital signature schemes \cite{elgamal1985public, johnson2001elliptic} in widely deployed protocols, including transport layer security \cite{blake2006elliptic}, secure shell \cite{stebila2009elliptic}, and in cryptocurrencies like Bitcoin \cite{blake1999standards, nakamoto2008peer}. 

In contrast to the computational hardness of ECDLP on classical computers, the security landscape changes fundamentally in the presence of large-scale quantum computers.
In 1994, Shor introduced a quantum algorithm that can solve integer factorization and discrete logarithm problems in polynomial time \cite{shor1994algorithms}.
Subsequent work showed that Shor’s algorithm can be generalized to elliptic curve groups \cite{proos2003shor}, implying that the ECDLP can also be efficiently solved on a fault-tolerant quantum computer and hence undermines the security of ECC.

At the same time, the practical implementation of quantum algorithms for cryptography problems remains severely constrained by the limitations of near-term quantum hardware. Fo instance, implementing Shor’s algorithm at cryptographically relevant scales requires a fault-tolerant quantum computer with a large number of logical qubits and circuits, as well as sufficiently low physical error rates (or equivalently achievable logical error rates after error correction) \cite{gidney2021factor, gidney2025factor}.
Improvements in one dimension often come at the cost of others \cite{roetteler2017quantum, haner2020improved}, making realistic resource optimization a delicate trade-off rather than a single-objective problem.

Motivated by these challenges, a substantial body of work has focused on reducing the quantum resource requirements of Shor’s algorithm for integer factorization \cite{regev2025efficient, ragavan2024space, kahanamoku2025jacobi, chevignard2025reducing}.
In particular, Chevignard et al.~\cite{chevignard2025reducing} demonstrated that RSA-2048 can be factored using approximately 1730 logical qubits.
By comparison, existing resource estimates indicate that solving the ECDLP on ECC-224, which offers a comparable classical security level, still requires at least 1862 logical qubits \cite{roetteler2017quantum, haner2020improved}. 

On the other hand, early systematic resource estimates for solving the ECDLP were provided by Roetteler et al.~\cite{roetteler2017quantum} and later refined in~\cite{haner2020improved}.
Subsequent studies investigated the impact of curve representations and field choices, with results exploring resource optimization on circuit depth~\cite{cryptoeprint:2026/106}, under specific hardware constraints~\cite{gu2025resource}, and several works suggesting that ECDLP over binary elliptic curves may admit more resource-efficient quantum implementations than prime-field curves \cite{banegas2020concrete, putranto2022another, jang2022optimized, taguchi2023concrete, jang2025new}.

These comparisons highlight that, despite extensive prior work, the quantum space requirements for solving the ECDLP remain relatively high, making the reduction of space usage a natural and important objective in the design of quantum ECDLP algorithms.

An important early theoretical result in this direction appeared in 2003, where Proos and Zalka \cite{proos2003shor} proposed an algorithmic approach via register sharing achieving an attractive asymptotic space bound of $5n + O(\sqrt{n})$ for solving the ECDLP over an $n$-bit prime field. Such a strategy was also recently adopted to qubit cost analysis among quantum algorithms for Decoded Quantum Interferometry (DQI)~\cite{khattar2025verifiable}. 
However, this result was derived at a high level of abstraction and did not include explicit quantum circuit constructions.
As a consequence, several issues critical for practical implementation were left unaddressed, including reversibility, ancilla management, and concrete resource estimation.
Moreover, the proposed approach incurs a fidelity loss of $O(n^{-1})$ arising from the approximate treatment of modular inversion, which further complicates its direct realization within the standard quantum circuit model.
As a result, it remains unclear to what extent the stated space bound can be achieved by a fully implementable quantum algorithm, leaving a practically relevant gap that has also been noted in subsequent resource estimation studies such as \cite{roetteler2017quantum}.

In conclusion, the motivation of our work can be summarized in one sentence:
\begin{center}
    {\em Can the quantum space requirements for solving the ECDLP be further reduced in practice?}
\end{center}

\input{sec_intro/sub_contribution}

%% file: sec_intro/sub_contribution.tex
\subsection*{Contributions}
Our main contribution is reducing the logical qubit requirements for solving ECDLP within the standard quantum circuit model. 
Building on the algorithmic framework of Proos and Zalka \cite{proos2003shor}, we present an explicit and fully reversible realization that achieves the previously established low-space bound at the circuit level, while remaining exactly correct, in contrast to the fidelity loss incurred in \cite{proos2003shor}. Specifically, our contributions are two-fold as follows.

\paragraph{Space-efficient, reversible algorithm design for modular inversion.}
We focus on modular inversion for computing $\ket{x}\ket{0} \to \ket{x}\ket{x^{-1} \bmod p}$ with $x \in \mathbb{F}_p$, which contributes the dominant component of the logical qubit requirement for solving ECDLP.
We develop a fully reversible algorithm for modular inversion based on the Extended Euclidean Algorithm (EEA), with details given in Section \ref{subsec:4phase}. 
Table \ref{tab:comp_qubit} and Table \ref{tab:comp_qubit_concrete} compare the logical qubit requirements for modular inversion in our work with those in previous studies.

\begin{table}[ht]
\centering
\small
\begin{tabular}{|c|c|c|}
    \hline
    \textbf{Source} & \textbf{Number of logical qubits} & \textbf{Remark} \\
    \hline
    \cite{proos2003shor} without register sharing & $5n + 4\log_2 n + O(1)$ & \multirow{2}{*}{without explicit circuit implementation} \\
    \cline{1-2}
    \cite{proos2003shor} with register sharing & $3n + 8\sqrt{n} + 4\log_2 n + O(1)$ & \\
    \hline
    \cite{roetteler2017quantum} & $7n + 2\log_2 n + O(1)$ & \multirow{3}{*}{with explicit circuit implementation}  \\
    \cline{1-2}
    \cite{haner2020improved} & $7n + \log_2 n + O(1)$ & \\
    \cline{1-2}
    \textbf{This work} & $3n + 4\log_2 n + O(1)$ & \\
    \hline
\end{tabular}
\caption{Comparison of the number of logical qubits required for modular inversion in ECDLP.}
\label{tab:comp_qubit}
\end{table}

\begin{table}[ht]
\centering
\small
\begin{tabular}{|c|cccccc|}
    \hline
    \multirow{2}{*}{\textbf{Source}} &
        \multicolumn{6}{c|}{\textbf{Number of logical qubits}} \\ \cline{2-7} 
     &
        \multicolumn{1}{c|}{ECC-160} &
        \multicolumn{1}{c|}{ECC-192} &
        \multicolumn{1}{c|}{ECC-224} &
        \multicolumn{1}{c|}{ECC-256} &
        \multicolumn{1}{c|}{ECC-384} &
        ECC-521 \\ \hline
    \cite{roetteler2017quantum} &
        \multicolumn{1}{c|}{1466} &
        \multicolumn{1}{c|}{1754} &
        \multicolumn{1}{c|}{2042} &
        \multicolumn{1}{c|}{2338} &
        \multicolumn{1}{c|}{3492} &
        4727 \\ \hline
    \cite{haner2020improved} &
        \multicolumn{1}{c|}{1350} &
        \multicolumn{1}{c|}{1606} &
        \multicolumn{1}{c|}{1862} &
        \multicolumn{1}{c|}{2124} &
        \multicolumn{1}{c|}{3151} &
        4258 \\ \hline
    \textbf{This work} &
        \multicolumn{1}{c|}{849} &
        \multicolumn{1}{c|}{1009} &
        \multicolumn{1}{c|}{1169} &
        \multicolumn{1}{c|}{1333} &
        \multicolumn{1}{c|}{1973} &
        2662 \\ \hline
\end{tabular}
\caption{Comparison of the concrete number of logical qubits required for Shor's algorithm.}
\label{tab:comp_qubit_concrete}
\end{table}

A central contribution of our work is a refined and explicit use of the register sharing strategy at the algorithmic level, as described in Section \ref{subsec:regshare}.
In the original construction of \cite{proos2003shor}, the same register resources are allocated to all input values $x \in \mathbb{F}_p$ in superposition.
This uniform allocation results in an $O(\sqrt{n})$ space overhead and requires truncating outlier values of $x$, which in turn introduces an $O(n^{-2})$ fidelity loss.
In contrast, we introduce an exact register allocation strategy that allows deterministic register usage across different input values $x \in \mathbb{F}_p$.
This strategy reduces the additional space overhead to $O(\log n)$ and ensures exact correctness for modular inversion.

Starting from a baseline reversible realization of our space-efficient Extended Euclidean Algorithm, we further introduce a sequence of algorithmic optimizations in Section \ref{subsec:opt}, that reorganize the update processes and remove redundant arithmetic operations.
To ensure correctness and reversibility, we provide complete algorithm pseudocode (Algorithm \ref{alg:opt}) together with concrete classical execution examples (Table \ref{tab:run_example}). 
These examples show that every step of the algorithm is exactly reversible and correct, and can be translated into a quantum circuit.

\paragraph{Explicit quantum circuit construction and resource estimation for modular inversion.}
To further show that our space-efficient algorithm for modular inversion is practically realizable, we present a detailed quantum circuit implementation in Section \ref{sec:circuit}. We begin by introducing the overall framework of the circuit implementation (shown in Figure \ref{fig:all_circuit_1} and Figure \ref{fig:all_circuit_2}). We then describe the detailed implementations of all building blocks  (shown in Figure \ref{fig:ctrl_swap} through Figure \ref{fig:len_update}).

We emphasize that these building blocks are non-standard components whose implementations are highly non-trivial. Consequently, the constructions presented here may not be optimal with respect to gate complexity or circuit depth.

\begin{figure}[htbp!]
    \centering
    \begin{subfigure}{0.48\linewidth}
        \centering
        \includegraphics[width=\linewidth]{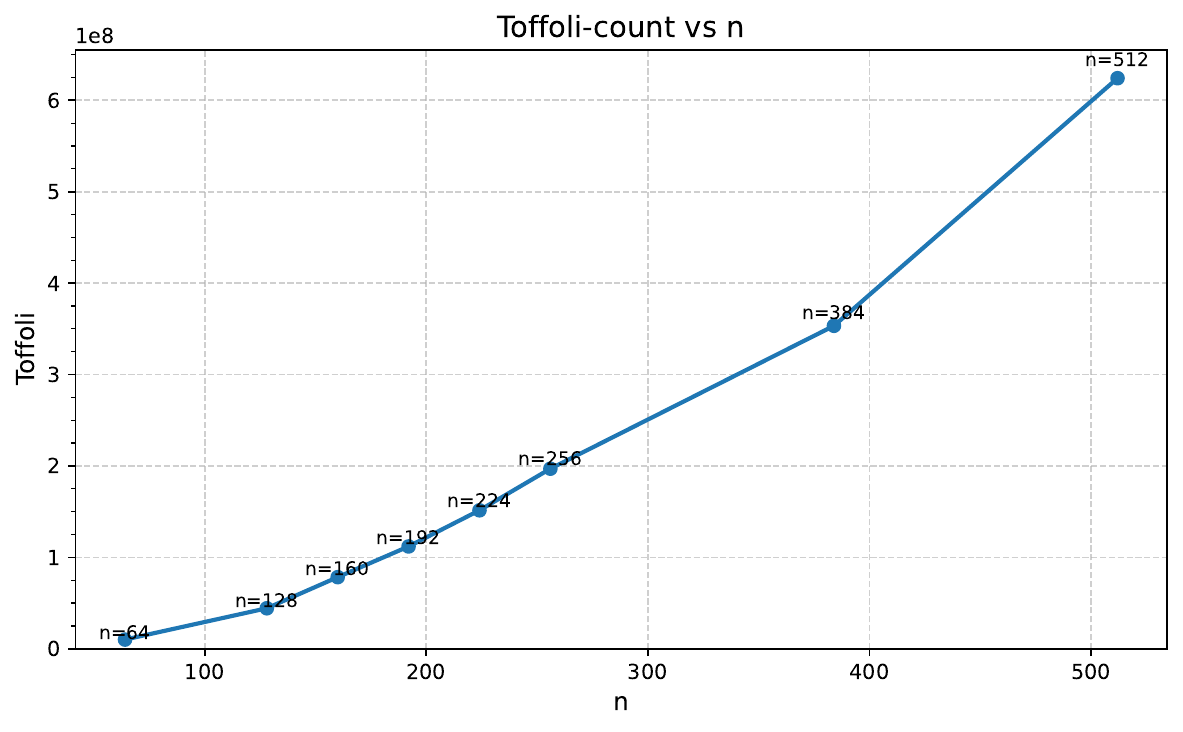}
        \caption{Toffoli-Count}
        \label{fig:toffoli}
    \end{subfigure}
    \hfill
    \begin{subfigure}{0.48\linewidth}
        \centering
        \includegraphics[width=\linewidth]{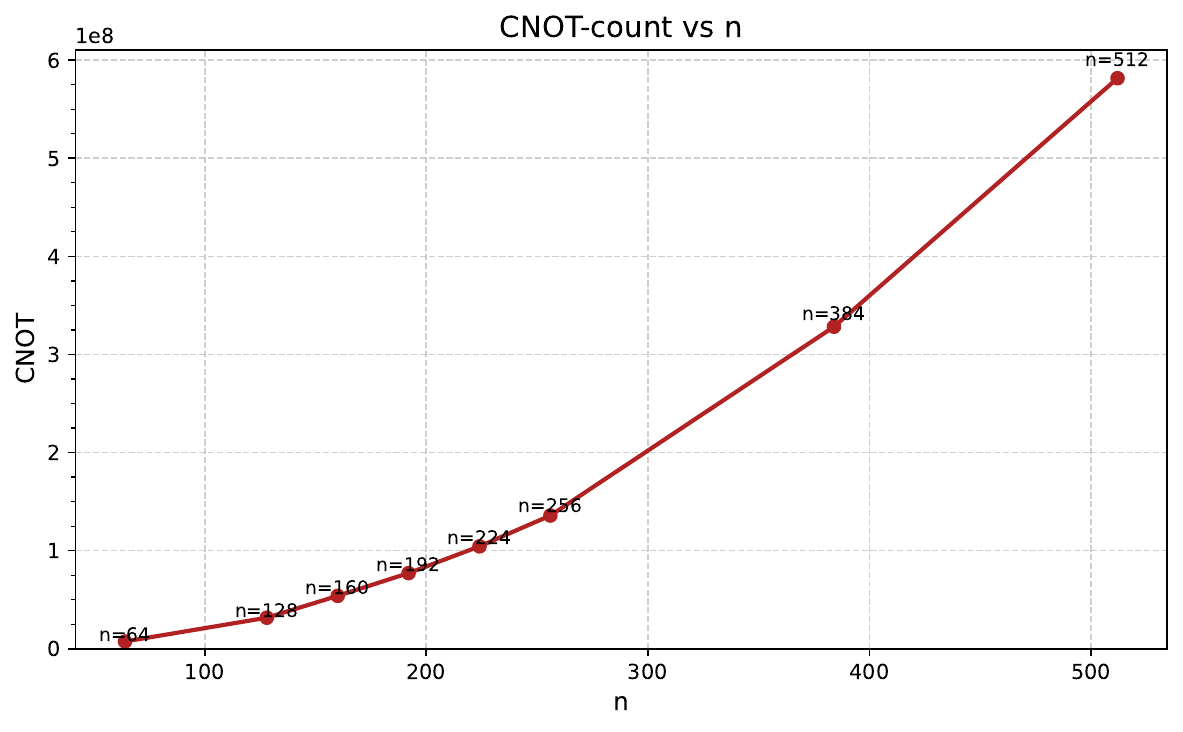}
        \caption{CNOT-count}
        \label{fig:cnot}
    \end{subfigure}
    
    \caption{Toffoli gate count and CNOT gate count for modular inversion in ECDLP.}
    \label{fig:toffli_counts}
\end{figure}

Finally, we provide resource estimates for our quantum circuit implementation. We show in Section \ref{subsec:inversion_costs} that one modular inversion requires approximately $204n^2\log_2 n + O(n^2)$ Toffoli gates and about $102n^2\log_2 n$ CNOT gates asymptotically. In addition, we present numerical evaluations of the Toffoli and CNOT gate counts as functions of the problem size $n$, as illustrated in Figure \ref{fig:toffli_counts}. We also merge our modular-inversion component into the full quantum circuit for solving the ECDLP and achieve an algorithm with  $5n + 4\lfloor \log_2 n \rfloor + O(1)$ qubits and $976n^3 + O\left(\frac{n^3}{\log_2 n}\right)$ Toffoli gates in total (see Section~\ref{subsec:component_costs}).

\paragraph{Open questions and concurrent work.} 
Our modular inversion circuit design serves as a general-purpose building block and is not limited to applications in elliptic curve cryptography. We expect that future work may further optimize its quantum gate counts and circuit depth, explore implementations and performance analysis on current quantum architectures, and extend this modular inversion block to other quantum algorithms.

During the preparation of this manuscript, Chevignard, Fouque, and Schrottenloher~\cite{cryptoeprint:2026/280} also proposed a quantum algorithm for ECDLP that reduces the number of qubits. Their improvement comes from the application of a residue number system to compute the projective coordinates of point multiplication, whereas our improvement comes from better implementation of the modular inversion by EEA. In terms of total Toffoli count, our algorithm achieves $O(n^3)$ while their algorithm is $O(n^{4}(\log n)^2)$.

Babbush et al.~\cite{babbush2026ECC} claimed implementation of Shor’s algorithm for ECDLP with significantly improved total Toffoli count. Because their technical details are not disclosed, we cannot make an explicit comparison with their work. In contrast, we present our implementation details as well as open-sourced repository (see Footnote~\ref{footnote:repo}).

%% file: sec_prelim/sec_prelim.tex
\section{Preliminaries}

This section provides a very brief discussion of the basic concepts used in this work. We assume that readers have basic knowledge of quantum computation. For a general survey of quantum computation, we recommend the survey written by Childs and van Dam~\cite{childs2010quantum}. 

\input{sec_prelim/sub_qc}
\input{sec_prelim/sub_ecdlp}
\input{sec_prelim/sub_shor}

%% file: sec_prelim/sub_qc.tex
\subsection{Quantum computing and the Toffoli networks}
\label{subsec:qc}

We write basis quantum states as $\ket{x}$, where $x$ is a bit string, and we model a quantum algorithm as a circuit composed of elementary gates applied to one or more qubits.
A key feature is that circuits can create and transform superpositions of basis states, enabling interference effects that are exploited by quantum algorithms.

At the fault-tolerant logical level it is common to express circuits over a universal gate set such as Clifford+$T$ \cite{nielsen2010quantum}.
In many architectures the $T$ gate is substantially more expensive than Clifford operations, so circuit cost is frequently summarized by $T$-count, $T$-depth, or related metrics \cite{fowler2012surface}.
For design and validation, however, it is often convenient to use a purely reversible gate basis.

Accordingly, we express the main reversible components as \emph{Toffoli networks}. The Toffoli gate maps
\[
    \ket{x,y,z}\ \longmapsto\ \ket{x,y, z \oplus (x y)},
\]
and it forms a universal primitive for classical reversible computation \cite{nielsen2010quantum}.
A practical benefit of working at the Toffoli-network level is that such networks admit exact realizations over Clifford+$T$ \cite{amy2013meet}, avoiding approximation overhead associated with synthesizing arbitrary unitaries.
Moreover, Toffoli-based reversible circuits can be efficiently simulated on classical inputs at scales far beyond what is possible for general quantum-state simulation, which makes testing and debugging large arithmetic circuits significantly more tractable \cite{haner2016factoring}.
For these reasons, we specify the core arithmetic and elliptic-curve routines using Toffoli and CNOT gates.

%% file: sec_prelim/sub_ecdlp.tex
\subsection{Elliptic curves and the ECDLP}
\label{subsec:ecdlp}

Let $p$ be a large prime and let $\mathbb{F}_p$ denote the finite field with $p$ elements.
An elliptic curve $E/\mathbb{F}_p$ (in short Weierstrass form) is the set of affine solutions $(x,y)\in \mathbb{F}_p^2$ to an equation $y^2 = x^3 + ax + b$, together with a distinguished point $\mathcal{O}$ at infinity.
The set of $\mathbb{F}_p$-rational points, together with $\mathcal{O}$, is denoted as $E(\mathbb{F}_p)$.

The points on $E(\mathbb{F}_p)$ form an abelian group under the elliptic-curve addition law, with identity element $\mathcal{O}$.
For affine points $P_1=(x_1,y_1)$ and $P_2=(x_2,y_2)$ that are not inverse to each other, the sum $P_3=P_1+P_2$ can be computed via a slope parameter $\lambda$: 
\[
    x_3 = \lambda^2 - x_1 - x_2, \qquad y_3 = \lambda(x_1 - x_3) - y_1,
\]
where $\lambda$ equals the chord slope $\frac{y_2 - y_1}{x_2 - x_1}$ when $P_1\neq P_2$, and the tangent slope $\frac{3x_1^2 + a}{2y_1}$ when $P_1 = P_2$.
Exceptional cases (e.g., $P_i = \mathcal{O}$ or $P_2 = -P_1$) are handled according the standard group properties.

For an integer $m\ge 1$, the notation $[m]P$ denotes the $m$-fold sum of a point $P$ with itself, i.e. $[m]P = P + P + \cdots + P$, where $P$ occurs $m$ time(s).
This $m$-fold sum can be extended to all $m\in \mathbb{Z}$ by defining $[0]P = \mathcal{O}$ and $[-m]P = [m](-P)$ for $m \ge 1$.
The operation $m\mapsto [m]P$ is called \emph{scalar multiplication} and is the core primitive used in elliptic-curve cryptography.

Let $P\in E(\mathbb{F}_p)$ generates a cyclic subgroup $\langle P\rangle$ of order $r$, and let $Q\in\langle P\rangle$.
The \emph{elliptic-curve discrete logarithm problem (ECDLP)} is to recover the unique $m\in\{0,1,\ldots,r-1\}$ such that
\[
    Q = [m]P.
\]
In classical settings, the best generic algorithms require on the order of $\Theta(\sqrt{p})$ group operations \cite{galbraith2016recent}, which is exponential in the bit length $\log_2 p$.

%% file: sec_prelim/sub_shor.tex
\subsubsection{Shor's quantum algorithm for ECDLP}
\label{subsec:shor}

Shor's discrete-logarithm algorithm for solving ECDLP applies to elliptic curve over any finite abelian group and therefore to $E(\mathbb{F}_p)$.
Let $n = \lfloor\log_2 p\rfloor + 1$ be the bit-length of $p$.
The quantum procedure uses two exponent registers and one register storing an elliptic curve point.

Firstly, initialize two $(n+1)$-qubit registers\footnote{Hasse's bound \cite{hasse1936theorie} indicates that $\mathrm{ord}(P) \le \#E(\F_p) \le p + 1 + 2\sqrt{p}$, which can be represented by at most $n + 1$ bits.} to be $\ket{0^{n+1}, 0^{n+1}}$, and apply Hadamard gate to all the qubits to obtain a uniform superposition over pairs $(k, \ell)\in \{0, 1, \ldots, 2^{n+1} - 1\}^2$.
Next, coherently compute the elliptic curve group action into an accumulator as
\begin{equation}\label{eqn:shor}
    \sum_{k, \ell = 0}^{2^{n+1} - 1}
    \ket{k, \ell}\ket{\mathcal{O}}
    \;\longmapsto\;
    \sum_{k, \ell = 0}^{2^{n+1} - 1}
    \ket{k, \ell}\ket{[k]P + [\ell]Q}.
\end{equation}
After this step, apply Quantum Fourier Transform $\QFT_{2^{n+1}}$ to both exponent registers and then measure them.
Finally, a classical post-processing procedure uses the measurement outcomes to reconstruct the discrete logarithm $m$ with high probability, as shown in \cite{shor1994algorithms}.

\begin{figure}[ht]
\centering
\footnotesize
\(\Qcircuit @C=0.6em @R=1.0em {
    & & & & & \ustick{\mu_0}\cwx[1] 
    & & & & &
    & & & \ustick{\mu_1}\cwx[1] & &
    & & & & &
    & & & & \ustick{\mu_{2n+1}}\cwx[1] \\
    & \lstick{\ket{0}} & \gate{H} & \ctrl{1} & \gate{H} & \meter 
    & & \ket{0} & & \gate{H} & \ctrl{1}
    & \gate{R_{\theta_1}} & \gate{H} & \meter & & \cdots
    & & & \ket{0} & & \gate{H}
    & \ctrl{1} & \gate{R_{\theta_{2n+1}}} & \gate{H} & \meter \\
    & \lstick{\ket{\mathcal{O}}} & \qw\slash^{2n} & \gate{+[2^0]P} & \qw & \qw
    & \qw & \qw & \qw & \qw & \gate{+[2^1]P} 
    & \qw & \qw & \qw & & \cdots
    & & & \qw & \qw & \qw 
    & \gate{+[2^n]Q} & \qw & \qw & \qw \\
}\)
\caption{The overall quantum circuit of Shor’s algorithm for solving ECDLP using semiclassical QFT. The gates $R_{\theta_i}$ denote rotation gates with rotation angle $\theta_i = \sum_{j = 0}^{i-1} 2^{i-j}\mu_j$, where the values $\mu_j\in \{0, 1\}$ are outcomes obtained in previous measurements. By employing the semiclassical QFT, the exponent register requires only a single qubit, resulting in a reduction of $2n+1$ qubits compared to the standard implementation.}
\label{fig:shor_semiQFT}
\end{figure}
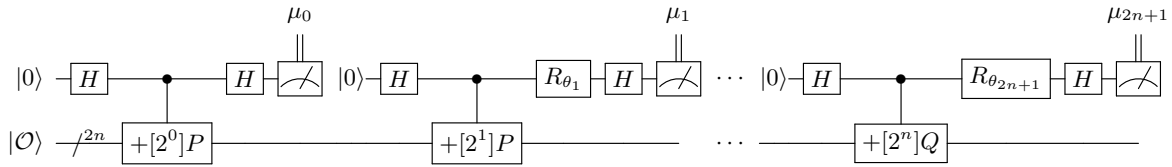

In terms of resource, the dominant cost arises from the coherent group evaluation, i.e., the double-scalar multiplication on elliptic curve points.
To reduce the qubit cost associated with QFT, its full circuit on the exponent registers can be replaced by a semiclassical variant \cite{griffiths1996semiclassical}.
This semiclassical Fourier transform performs measurements during the computation, reusing qubits and applying conditional phase rotations based on previously observed outcomes.
The overall quantum circuits of Shor's algorithm for ECDLP using QFT (resp. semiclassical QFT) are shown in Figure \ref{fig:shor_QFT} (resp. Figure \ref{fig:shor_semiQFT}).
\begin{figure}[ht]
\centering
\footnotesize
\(\Qcircuit @C=0.8em @R=0.8em {
    & \lstick{\ket{0}} & \gate{H} & \ctrl{10} & \qw & \qw
    & \cdots & & \qw & \qw & \qw 
    & \qw & \cdots & & \qw & \qw 
    & \multigate{4}{\QFT_{2^{n+1}}} & \meter \\
    & \lstick{\ket{0}} & \gate{H} & \qw & \ctrl{9} & \qw 
    & \cdots & & \qw & \qw & \qw 
    & \qw & \cdots & & \qw & \qw 
    & \ghost{\QFT_{2^{n+1}}} & \meter \\
    & \lstick{\vdots\ } & & & & 
    & \vdots & & & &
    & & \vdots & & & 
    & & \\
    & & & & & & & & & & & & & & & & & \\
    & \lstick{\ket{0}} & \gate{H} & \qw & \qw & \qw 
    & \cdots & & \ctrl{6} & \qw & \qw 
    & \qw & \cdots & & \qw & \qw 
    & \ghost{\QFT_{2^{n+1}}} & \meter \\
    & \lstick{\ket{0}} & \gate{H} & \qw & \qw & \qw 
    & \cdots & & \qw & \ctrl{5} & \qw 
    & \qw & \cdots & & \qw & \qw 
    & \multigate{4}{\QFT_{2^{n+1}}} & \meter \\
    & \lstick{\ket{0}} & \gate{H} & \qw & \qw & \qw 
    & \cdots & & \qw & \qw & \ctrl{4} 
    & \qw & \cdots & & \qw & \qw 
    & \ghost{\QFT_{2^{n+1}}} & \meter \\
    & \lstick{\vdots\ } & & & & 
    & \vdots & & & &
    & & \vdots & & & 
    & & \\
    & & & & & & & & & & & & & & & & \\
    & \lstick{\ket{0}} & \gate{H} & \qw & \qw & \qw 
    & \cdots & & \qw & \qw & \qw 
    & \qw & \cdots & & \ctrl{1} & \qw 
    & \ghost{\QFT_{2^{n+1}}} & \meter \\
    & \lstick{\ket{\mathcal{O}}} & \qw\slash^{2n} & \gate{+[2^0]P} & \gate{+[2^1]P} & \qw 
    & \cdots & & \gate{+[2^n]P} & \gate{+[2^0]Q} & \gate{+[2^1]Q}
    & \qw & \cdots & & \gate{+[2^n]Q} & \qw 
    & \qw & \qw \\
}\)
\caption{The overall quantum circuit of Shor’s algorithm for solving ECDLP using QFT. The qubits, from top to bottom, correspond to the exponent registers containing $k$ and $\ell$ in Equation \ref{eqn:shor} (ordered from lower-order bits to higher-order bits), and to the register that stores the elliptic curve point accumulator.
}
\label{fig:shor_QFT}
\end{figure}
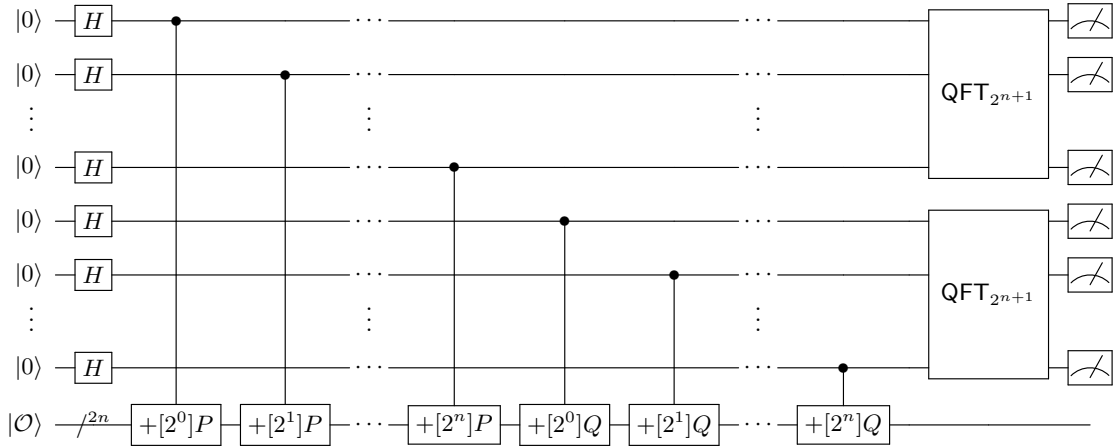

%% file: sec_EEA/sec_EEA.tex
\section{Space-Efficient Extended Euclidean Algorithm}
\label{sec:EEA}

As discussed in previous sections, the dominant cost in designing quantum circuits for solving ECDLP via Shor’s algorithm arises from coherent group additions on elliptic curve points. Among these operations, modular inversion is the most resource-consuming component. In this section, we present a space-efficient and reversible algorithm for modular inversion based on the Extended Euclidean Algorithm. This design is suitable for further translation into a quantum circuit, as detailed in Section \ref{sec:circuit}.

Suppose that $p > x$ are two positive integers. The well-known \emph{Euclidean algorithm} can be used to compute the greatest common divisor (GCD) of $p$ and $x$. The algorithm initializes $r_0 := p$ and $r_1 := x$, and for iterations $i = 1, 2, \ldots$, it repeatedly divides $r_{i-1}$ by $r_i$ to obtain the quotient $q_i = \lfloor r_{i-1} / r_i \rfloor$ and the remainder $r_{i+1} = r_{i-1} - q_i r_i$. The process terminates when $r_{k+1} = 0$, at which point the greatest common divisor of $p$ and $x$ equals $r_k$.

The \emph{Extended Euclidean Algorithm (EEA)} not only computes the GCD of $p$ and $x$, but also outputs the modular inverse $x^{-1} \bmod p$ when $p$ and $x$ are coprime. In addition to computing the sequence $r_i$, it also initializes $t_0 := 0$ and $t_1 := 1$, and for each iteration $i = 1, 2, \ldots$, updates $t_{i+1} = t_{i-1} + q_i t_i$ after obtaining $q_i = \lfloor r_{i-1} / r_i \rfloor$. When the algorithm first encounters $r_k = 1$, the modular inverse is given by $(-1)^{k-1} t_k \bmod p$.

When realizing the Extended Euclidean algorithm on a quantum computer to compute the modular inverse of $x$ with a prime modulus $p$ of $n$ binary digits, the primary challenge lies in handling superpositions of $x$. The number of iterations required varies from $1$ to $O(n)$ depending on the input, implying that a straightforward implementation would require $O(n)$ iterations. Each iteration involves divisions and multiplications on $n$-digit numbers, leading to an overall gate complexity of $O(n^3)$.

In what follows, we describe the dedicated four-phase algorithm framework first proposed in \cite{proos2003shor}, followed by our own algorithmic optimizations and the details of our optimized circuit implementation.

\input{sec_EEA/sub_4phase}
\input{sec_EEA/sub_regshare}
\input{sec_EEA/sub_opt}

%% file: sec_EEA/sub_4phase.tex
\subsection{The four-phase algorithm framework}
\label{subsec:4phase}

The four-phase algorithm framework was originally introduced in \cite{proos2003shor}. It operates on six quantum registers, denoted as $r_{i-1}$, $r_i$, $t_{i-1}$, $t_i$, $q_i$, and $\ell$, where the first five registers correspond to the variables used in the $i$-th iteration of EEA. To compute the quotient $q_i = \lfloor r_{i-1} / r_i \rfloor$, the algorithm performs a binary long-division procedure consisting solely of bit shifts and long additions or subtractions. The algorithm proceeds through the following four phases:
\begin{itemize}
    \item \textbf{Phase 1.} The register containing $r_i$ are repeatedly shifted left by one bit, while the shift counter $\ell$ is incremented at each step. This continues until the inequality $2^\ell r_i > r_{i-1}$ holds. Conceptually, this phase determines the largest power of two by which $r_i$ can be multiplied without exceeding $r_{i-1}$. Equivalently, it identifies the bit-length of the binary representation of $q_i$, which we denote by $\ell_0$.

    \item \textbf{Phase 2.} The register containing $r_i$ are then shifted right by one bit per step, with $\ell$ decremented accordingly. At each step, the registers corresponding to $r_{i-1}$ and $q_i$ are updated as
    \[
        (r_{i-1} - q' r_i, q') 
        \to \left(r_{i-1} - (q' + 2^\ell q_{i, \ell}) r_i, q' + 2^\ell q_{i, \ell}\right),
    \]
    depending on whether $r_{i-1} - q' r_i$ remains greater than $2^{\ell} r_i$. Here $q'$ denotes the current partial quotient, given by $q' = 2^{\ell_0} q_{i, \ell_0} + 2^{\ell_0 - 1} q_{i, \ell_0 - 1} + \cdots + 2^{\ell + 1} q_{i, \ell + 1}$, representing the most significant bits accumulated so far in the binary expansion
    \[
        q_i = 2^{\ell_0} q_{i, \ell_0} + 2^{\ell_0 - 1} q_{i, \ell_0 - 1} + \cdots + q_{i, 0}.
    \]
    This iterative process ultimately yields the transformation
    \[
        (r_{i-1}, 0) 
        \to (r_{i-1} - q_i r_i, q_i),
    \]
    completing the computation of $q_i$.

    \item \textbf{Phase 3.} The register containing $t_i$ are shifted left by one bit per step, and $\ell$ is incremented correspondingly. The registers associated with $t_{i-1}$ and $q_i$ are updated as
    \[
        (t_{i-1} + q'' t_i, q') 
        \to \left(t_{i-1} + (q'' + 2^\ell q_{i, \ell}) t_i, q'\right),
    \]
    where the update depends on the bit value $q_{i, \ell}$ of $q_i$. After each step, $q_{i, \ell}$ is reset to $0$, depending on whether $t_{i-1} + (q'' + 2^\ell q_{i, \ell}) t_i > 2^{\ell} t_i$. Here $q'$ and $q''$ represent the most and least significant portions of the quotient, respectively:
    \[
        q' = 2^{\ell_0} q_{i, \ell_0} + 2^{\ell_0 - 1} q_{i, \ell_0 - 1} + \cdots + 2^{\ell} q_{i, \ell}, \quad
        q'' = 2^{\ell - 1} q_{i, \ell - 1} + 2^{\ell - 2} q_{i, \ell - 2} + \cdots + q_{i, 0}.
    \]
    After completing all steps, the registers corresponding to $t_{i-1}$ and $q_i$ are updated as
    \[
        (t_{i-1}, q_i) 
        \to (t_{i-1} + q_i t_i, 0).
    \]

    \item \textbf{Phase 4.} Finally, the register containing $t_i$ are repeatedly shifted right by one bit, and $\ell$ is decremented at each step. The process terminates once $\ell = 0$.
\end{itemize}

Upon completion of all four phases, a SWAP gate is applied between the register pairs corresponding to $(r_{i-1}, r_i)$ (which is currently $(r_{i+1}, r_i)$) and the register pairs corresponding to  $(t_{i-1}, t_i)$ (which is currently $(t_{i+1}, t_i)$), respectively. This completes one full iteration of the EEA implementation.

We emphasize that, during the execution of this four-phase algorithm, the correspondence between the iteration index in EEA and the phase number may vary depending on the input value $x$. In other words, the algorithm’s internal progression through iterations and phases is input-dependent. To avoid ambiguity in the subsequent descriptions, we use ``step'' to represent any sub-iteration occurring within any phase of an EEA iteration. The example in Table \ref{tab:step_status} illustrates how the step status evolves for different input values under this four-phase framework.
\begin{table}[htbp]
\centering
\footnotesize
\begin{tabular}{cccc}
\hline
\textbf{Input}  & $x_1$ & $x_2$ & $x_3$ \\
\hline
\textbf{Step 1} & Iteration 1, Phase 1 & Iteration 1, Phase 1      & Iteration 1, Phase 1 \\
\hline
\textbf{Step 2} & Iteration 1, Phase 1 & Iteration 1, Phase 1      & Iteration 1, Phase 1 \\
\hline
\textbf{Step 3} & Iteration 1, Phase 2 & Iteration 1, Phase 1      & Iteration 1, Phase 1 \\
\hline
\textbf{Step 4} & Iteration 1, Phase 2 & Iteration 1, Phase 1      & Iteration 1, Phase 2 \\
\hline
\textbf{Step 5} & Iteration 1, Phase 3 & Iteration 1, Phase 1      & Iteration 1, Phase 2 \\
\hline
\textbf{Step 6} & Iteration 1, Phase 3 & Iteration 1, Phase 2      & Iteration 1, Phase 2 \\
\hline
\textbf{Step 7} & Iteration 1, Phase 4 & Iteration 1, Phase 2      & Iteration 1, Phase 3 \\
\hline
\textbf{Step 8} & Iteration 1, Phase 4 (and SWAP) & Iteration 1, Phase 2      & Iteration 1, Phase 3 \\
\hline
\textbf{Step 9} & Iteration 2, Phase 1 & Iteration 1, Phase 2      & Iteration 1, Phase 3 \\
\hline
\textbf{Step 10} & Iteration 2, Phase 1 & Iteration 1, Phase 2      & Iteration 1, Phase 4 \\
\hline
\textbf{Step 11} & Iteration 2, Phase 1 & Iteration 1, Phase 3      & Iteration 1, Phase 4 \\
\hline
\textbf{Step 12} & Iteration 2, Phase 2 & Iteration 1, Phase 3      & Iteration 1, Phase 4 (and SWAP) \\
\hline
\textbf{Step 13} & Iteration 2, Phase 2 & Iteration 1, Phase 3      & Iteration 2, Phase 1 \\
\hline
\textbf{Step 14} & Iteration 2, Phase 2 & Iteration 1, Phase 3      & Iteration 2, Phase 1 \\
\hline
\textbf{Step 15} & Iteration 2, Phase 3 & Iteration 1, Phase 3      & Iteration 2, Phase 2 \\
\hline
\textbf{Step 16} & Iteration 2, Phase 3 & Iteration 1, Phase 4      & Iteration 2, Phase 2 \\
\hline
\textbf{$\cdots$} & $\cdots$ & $\cdots$ & $\cdots$ \\
\hline
\textbf{Step 100} & Iteration 9, Phase 3 & Iteration 5, Phase 2      & Iteration 8, Phase 4 \\
\hline
\textbf{$\cdots$} & $\cdots$ & $\cdots$ & $\cdots$ \\
\hline
\end{tabular}
\caption{An example illustrating that the iteration index in EEA. The phase number may vary depending on the input value $x$.}
\label{tab:step_status}
\end{table}

Although the sequence of step statuses differs for each input $x$, the total number of steps required to reach an index $k$ such that $r_{k-1} = 1, r_k = 0$ can be bounded within the interval $[4n, 4\lceil cn\rceil]$ with $c = 1 / \log_2\left(\frac{\sqrt{5} + 1}{2}\right)$, as formally proven in Appendix \ref{app:step_num}. This bounded-step property is crucial for quantum implementation: it implies that a quantum circuit corresponding to one ``step'' can be repeated at most $4\lceil cn\rceil \approx 5.76n$ times to definitely obtain the modular inverse of any input $x$ in superposition. Since each step involves arithmetic operations such as shifts, additions, and subtractions, which require $O(n)$ quantum gates, the overall gate complexity of the algorithm is reduced to $O(n^2)$. 

Keeping the above discussion in mind, the complete space-efficient EEA for computing the modular inverse $x^{-1} \bmod p$, where $p$ is a prime represented with $n$ binary digits and $x \in \{1, 2, \ldots, p - 1\}$, is summarized in Algorithm \ref{alg:full_EEA}. In this context: (i) the register allocation strategy is described in Section \ref{subsec:regshare}; (ii) Algorithm \ref{alg:opt} provides the optimized stepwise iteration designed to minimize gate complexity and ensure quantum reversibility, as further discussed in Sections \ref{subsec:regshare} and \ref{subsec:opt}; (iii) the preliminary operation of replacing $x$ by $p - x$ when $x > p/2$ guarantees that the entire procedure remains logically reversible under quantum computation.
\begin{algorithm}[htbp]
\small
\caption{Full description of our space-efficient EEA}
\begin{algorithmic}
\Require Quantum registers $\ket{x}\ket{0^n}\ket{0^m}$, where the first $2n$ qubits hold the input/output data and the remaining $m$ serve as auxiliary qubits.
\Ensure Output state $\ket{x}\ket{x^{-1}\bmod p}\ket{0^m}$.

\State Initialize register \texttt{Work1} with $(n + 3)$ qubits as $\ket{100, p}$ \Comment{Using $(n + 3)$ auxiliary qubits}
\State Initialize register \texttt{Work2} with $(n + 3)$ qubits as $\ket{000, x}$ \Comment{Pad $x$ to $n$ bits on the left}
\State Initialize \texttt{Length} registers to $\ket{\ell_t := 1, \ell_q := 0, \ell_{r'} := \text{length of } x, \ell_s := 0}$
\State Initialize \texttt{Control} registers to $\ket{\texttt{Phase1} := 0, \texttt{Phase2} := 0, \texttt{Sign} := 0, \texttt{Iter} := 0}$

\If{$x > p / 2$} \Comment{Ensures that the algorithm is invertible}
    \State $\texttt{Iter} \gets \texttt{Iter} \oplus 1$
\EndIf
\If{$\texttt{Iter} = 1$}
    \State $x \gets p - x$
\EndIf

\For{$i = 1, 2, \cdots, 4\lceil cn\rceil$} \Comment{Main iterative loop performing the optimized algorithm steps}
    \State Execute Algorithm~\ref{alg:opt} on the quantum registers
\EndFor

\State Copy the content of $t'$ in \texttt{Work2} to the output register
\If{$\texttt{Iter} = 0$} \Comment{The resulting value equals $(-1)^k t_k$, where $k$ is the number of EEA iterations}
    \State $\texttt{Output} \gets p - \texttt{Output}$
\EndIf

\For{$i = 1, 2, \cdots, 4\lceil cn\rceil$} \Comment{Reverse computation to recompute $x$}
    \State Execute the inverse of Algorithm~\ref{alg:opt} on the quantum registers
\EndFor

\If{$\texttt{Iter} = 1$} \Comment{Undo the earlier transformation $x \gets p - x$}
    \State $x \gets p - x$
\EndIf
\If{$x > p / 2$}
    \State $\texttt{Iter} \gets \texttt{Iter} \oplus 1$
\EndIf

\State Reset all auxiliary qubits to $\ket{0}$.

\end{algorithmic}
\label{alg:full_EEA}
\end{algorithm}

%% file: sec_EEA/sub_regshare.tex
\subsection{Space-efficient algorithm by register sharing}
\label{subsec:regshare}

In \cite{proos2003shor}, the authors introduced the concept of \emph{register sharing}. The central idea of register sharing is that, during the execution of the EEA, the sequence $\{r_i\}$ is monotonically decreasing while the sequence $\{t_i\}$ is monotonically increasing. This complementary behavior allows both values to occupy the same quantum register at different stages of computation, thereby reducing the overall space requirement. The identity
\[
    r_{i-1}t_i + r_it_{i-1} = p
\]
implies that two quantum registers, each consisting of $(n + 2)$ qubits, are sufficient: one to store the pair $(r_{i-1}, t_i)$ and the other to store $(r_i, t_{i-1})$.

\begin{figure}[ht]
\centering

\begin{tikzpicture}[
    scale=0.5,
    transform shape,
    font=\Huge,
    >=Stealth,
    segR/.style={draw=myred, line width=6pt, line cap=rect},
    segB/.style={draw=myblue, line width=6pt, line cap=rect},
    segG/.style={draw=mygreen, line width=6pt, line cap=rect},
    braceR1/.style={decorate, decoration={brace, amplitude=4pt, mirror}, draw=myred, line width=1pt},
    braceR2/.style={decorate, decoration={brace, amplitude=4pt}, draw=myred, line width=1pt},
    braceB/.style={decorate, decoration={brace, amplitude=4pt}, draw=myblue, line width=1pt},
    braceG/.style={decorate, decoration={brace, amplitude=4pt, mirror}, draw=mygreen, line width=1pt},
    phaseArrow/.style={->, line width=1pt},
]

\usetikzlibrary{decorations.pathreplacing,arrows.meta,positioning,calc}
\usetikzlibrary{arrows.meta,decorations.pathreplacing,calc}

\definecolor{myred}{RGB}{240,40,40}
\definecolor{myblue}{RGB}{40,40,240}
\definecolor{mygreen}{RGB}{0,160,0}

\def\Xl{0.0}
\def\Xre{10.0}
\def\Ytop{0.0}
\def\Ybot{-1.8}
\def\rDist{1.4}

\def\Lt{4.0}
\def\Ltp{3.0}
\def\Lr{5.0}
\def\Lrp{3.0}
\def\Lrs{2.5}
\def\Lq{1.5}
\def\Ls{1.4}
\def\Gap{0.6}

\pgfmathsetmacro{\Xr}{\Xre-\Lr}     
\pgfmathsetmacro{\Xrp}{\Xre-\Lrp}   
\pgfmathsetmacro{\Xrs}{\Xre-\Lrs}   

\coordinate (A) at (0,0);        
\coordinate (B) at (16,0);      
\coordinate (C) at (16,-6);   
\coordinate (D) at (0,-6);    
\coordinate (E) at (0,-12);      
\coordinate (F) at (16,-12);    

\begin{scope}[shift={(A)}]
    \draw[line width=1.0pt] (\Xl-\rDist,\Ytop+\rDist) rectangle (\Xre+\rDist,\Ybot-\rDist);
    
    \coordinate (tA1) at (\Xl,\Ytop);
    \coordinate (tA2) at (\Xl+\Lt,\Ytop);
    \draw[segR] (tA1) -- (tA2);
    \node[above=6pt, text=myred] at ($(tA1)!0.5!(tA2)$) {$t$};
    \draw[braceR1] (\Xl,\Ytop-0.3) -- (\Xl+\Lt,\Ytop-0.3) node[midway, below=6pt, text=myred] {\huge{$\ell_t$}};
    
    \coordinate (rA1) at (\Xr,\Ytop);
    \coordinate (rA2) at (\Xre,\Ytop);
    \draw[segB] (rA1) -- (rA2);
    \node[above=6pt, text=myblue] at ($(rA1)!0.5!(rA2)$) {$r$};
    
    \coordinate (tpA1) at (\Xl,\Ybot);
    \coordinate (tpA2) at (\Xl+\Ltp,\Ybot);
    \draw[segR] (tpA1) -- (tpA2);
    \node[below=6pt, text=myred] at ($(tpA1)!0.5!(tpA2)$) {$t'$};
    
    \coordinate (rpA1) at (\Xrp,\Ybot);
    \coordinate (rpA2) at (\Xre,\Ybot);
    \draw[segB] (rpA1) -- (rpA2);
    \node[below=6pt, text=myblue] at ($(rpA1)!0.5!(rpA2)$) {$r'$};
    \draw[braceB] (\Xrp,\Ybot+0.3) -- (\Xre,\Ybot+0.3) node[midway, above=6pt, text=myblue] {\huge{$\ell_{r'}$}};
\end{scope}

\draw[phaseArrow] ($(A)+(11.7,-0.9)$) -- ($(B)+(-1.7,-0.9)$)
node[midway, above=4pt] {\huge{\texttt{Phase 1}}};

\begin{scope}[shift={(B)}]
    \draw[line width=1.0pt] (\Xl-\rDist,\Ytop+\rDist) rectangle (\Xre+\rDist,\Ybot-\rDist);
    
    \coordinate (tB1) at (\Xl,\Ytop);
    \coordinate (tB2) at (\Xl+\Lt,\Ytop);
    \draw[segR] (tB1) -- (tB2);
    \node[above=6pt, text=myred] at ($(tB1)!0.5!(tB2)$) {$t$};
    
    \coordinate (rB1) at (\Xr,\Ytop);
    \coordinate (rB2) at (\Xre,\Ytop);
    \draw[segB] (rB1) -- (rB2);
    \node[above=6pt, text=myblue] at ($(rB1)!0.5!(rB2)$) {$r$};
    
    \coordinate (sL1) at (\Xl,\Ybot);
    \coordinate (sL2) at (\Xl+\Ltp-\Ls,\Ybot);
    \draw[segR] (sL1) -- (sL2);
    \node[below=6pt, text=myred] at ($(sL1)!0.5!(sL2)$) {$t'$(h)};

    \coordinate (rpB1) at (\Xr,\Ybot);               
    \coordinate (rpB2) at (\Xre-\Ls-\Gap,\Ybot); 
    \draw[segB] (rpB1) -- (rpB2);
    \node[below=6pt, text=myblue] at ($(rpB1)!0.5!(rpB2)$) {$r'$};

    \coordinate (sR1) at (\Xre-\Ls,\Ybot);
    \coordinate (sR2) at (\Xre,\Ybot);
    \draw[segR] (sR1) -- (sR2);
    \node[below=6pt, text=myred] at ($(sR1)!0.5!(sR2)$) {$t'$(l)};
    \draw[braceR2] (\Xre-\Ls,\Ybot+0.3) -- (\Xre,\Ybot+0.3) node[midway, above=6pt, text=myred] {\huge{$\ell_s$}};
\end{scope}

\draw[phaseArrow] ($(B)+(5.0,-3.4)$) -- ($(C)+(5.0,1.6)$);
\node[anchor=east] at ($(B)+(4.8,-3.8)$) {\huge{\texttt{Phase 2}}};

\begin{scope}[shift={(C)}]
    \draw[line width=1.0pt] (\Xl-\rDist,\Ytop+\rDist) rectangle (\Xre+\rDist,\Ybot-\rDist);
    
    \coordinate (tC1) at (\Xl,\Ytop);
    \coordinate (tC2) at (\Xl+\Lt,\Ytop);
    \draw[segR] (tC1) -- (tC2);
    \node[above=6pt, text=myred] at ($(tC1)!0.5!(tC2)$) {$t$};

    \coordinate (rsC1) at (\Xrs,\Ytop);
    \coordinate (rsC2) at (\Xre,\Ytop);
    \draw[segB] (rsC1) -- (rsC2);
    \node[above=6pt, text=myblue] at ($(rsC1)!0.5!(rsC2)$) {$r-qr'$};

    \coordinate (qC1) at (\Xl+\Lt+\Gap,\Ytop);
    \coordinate (qC2) at (\Xl+\Lt+\Gap+\Lq,\Ytop);
    \draw[segG] (qC1) -- (qC2);
    \node[above=6pt, text=mygreen] at ($(qC1)!0.5!(qC2)$) {$q$};
    \draw[braceG] ($(qC1)+(0,-0.3)$) -- ($(qC2)+(0,-0.3)$) node[midway, below=6pt, text=mygreen] {\huge{$\ell_q$}};

    \coordinate (tpC1) at (\Xl,\Ybot);
    \coordinate (tpC2) at (\Xl+\Ltp,\Ybot);
    \draw[segR] (tpC1) -- (tpC2);
    \node[below=6pt, text=myred] at ($(tpC1)!0.5!(tpC2)$) {$t'$};

    \coordinate (rpC1) at (\Xrp,\Ybot);
    \coordinate (rpC2) at (\Xre,\Ybot);
    \draw[segB] (rpC1) -- (rpC2);
    \node[below=6pt, text=myblue] at ($(rpC1)!0.5!(rpC2)$) {$r'$};
\end{scope}

\draw[phaseArrow] ($(C)+(-1.7,-0.9)$) -- ($(D)+(11.7,-0.9)$)
node[midway, above=4pt] {\huge{\texttt{Phase 3}}};

\begin{scope}[shift={(D)}]
    \draw[line width=1.0pt] (\Xl-\rDist,\Ytop+\rDist) rectangle (\Xre+\rDist,\Ybot-\rDist);

    \coordinate (tD1) at (\Xl,\Ytop);
    \coordinate (tD2) at (\Xl+\Lt,\Ytop);
    \draw[segR] (tD1) -- (tD2);
    \node[above=6pt, text=myred] at ($(tD1)!0.5!(tD2)$) {$t$};

    \coordinate (rsD1) at (\Xrs,\Ytop);
    \coordinate (rsD2) at (\Xre,\Ytop);
    \draw[segB] (rsD1) -- (rsD2);
    \node[above=6pt, text=myblue] at ($(rsD1)!0.5!(rsD2)$) {$r-qr'$};

    \coordinate (lD1) at (\Xl,\Ybot);
    \coordinate (lD2) at (\Xre-\Ls-\Gap-\Lrp-1.0,\Ybot);
    \draw[segR] (lD1) -- (lD2);
    \node[below=6pt, text=myred] at ($(lD1)!0.5!(lD2)$) {$t'$(h)};

    \coordinate (rpD1) at (\Xre-\Ls-\Gap-\Lrp,\Ybot);     
    \coordinate (rpD2) at (\Xre-\Ls-\Gap,\Ybot);
    \draw[segB] (rpD1) -- (rpD2);
    \node[below=6pt, text=myblue] at ($(rpD1)!0.5!(rpD2)$) {$r'$};

    \coordinate (sD1) at (\Xre-\Ls,\Ybot);
    \coordinate (sD2) at (\Xre,\Ybot);
    \draw[segR] (sD1) -- (sD2);
    \node[below=6pt, text=myred] at ($(sD1)!0.5!(sD2)$) {$t'$(l)};
\end{scope}

\draw[phaseArrow] ($(D)+(5.0,-3.4)$) -- ($(E)+(5.0,1.6)$);
\node[anchor=west] at ($(D)+(5.2,-3.8)$) {\huge{\texttt{Phase 4}}};

\begin{scope}[shift={(E)}]
    \draw[line width=1.0pt] (\Xl-\rDist,\Ytop+\rDist) rectangle (\Xre+\rDist,\Ybot-\rDist);

    \coordinate (tE1) at (\Xl,\Ytop);
    \coordinate (tE2) at (\Xl+\Lt,\Ytop);
    \draw[segR] (tE1) -- (tE2);
    \node[above=6pt, text=myred] at ($(tE1)!0.5!(tE2)$) {$t$};

    \coordinate (rsE1) at (\Xrs,\Ytop);
    \coordinate (rsE2) at (\Xre,\Ytop);
    \draw[segB] (rsE1) -- (rsE2);
    \node[above=6pt, text=myblue] at ($(rsE1)!0.5!(rsE2)$) {$r-qr'$};

    \coordinate (bigE1) at (\Xl,\Ybot);
    \coordinate (bigE2) at (\Xrp-1.0,\Ybot);
    \draw[segR] (bigE1) -- (bigE2);
    \node[below=6pt, text=myred] at ($(bigE1)!0.5!(bigE2)$) {$t'+qt$};

    \coordinate (rpE1) at (\Xrp,\Ybot);
    \coordinate (rpE2) at (\Xre,\Ybot);
    \draw[segB] (rpE1) -- (rpE2);
    \node[below=6pt, text=myblue] at ($(rpE1)!0.5!(rpE2)$) {$r'$};
\end{scope}

\draw[phaseArrow] ($(E)+(11.7,-0.9)$) -- ($(F)+(-1.7,-0.9)$)
node[midway, above=4pt] {\huge{\texttt{Swap}}};

\begin{scope}[shift={(F)}]
    \draw[line width=1.0pt] (\Xl-\rDist,\Ytop+\rDist) rectangle (\Xre+\rDist,\Ybot-\rDist);

    \coordinate (bigF1) at (\Xl,\Ytop);
    \coordinate (bigF2) at (\Xrp-1.0,\Ytop);
    \draw[segR] (bigF1) -- (bigF2);
    \node[above=6pt, text=myred] at ($(bigF1)!0.5!(bigF2)$) {$t'+qt$};

    \coordinate (rpF1) at (\Xrp,\Ytop);
    \coordinate (rpF2) at (\Xre,\Ytop);
    \draw[segB] (rpF1) -- (rpF2);
    \node[above=6pt, text=myblue] at ($(rpF1)!0.5!(rpF2)$) {$r'$};

    \coordinate (tF1) at (\Xl,\Ybot);
    \coordinate (tF2) at (\Xl+\Lt,\Ybot);
    \draw[segR] (tF1) -- (tF2);
    \node[below=6pt, text=myred] at ($(tF1)!0.5!(tF2)$) {$t$};

    \coordinate (rsF1) at (\Xrs,\Ybot);
    \coordinate (rsF2) at (\Xre,\Ybot);
    \draw[segB] (rsF1) -- (rsF2);
    \node[below=6pt, text=myblue] at ($(rsF1)!0.5!(rsF2)$) {$r-qr'$};
\end{scope}
\end{tikzpicture}

\caption{An illustration of how the two \texttt{Work} registers are allocated for temporary variables within a single iteration of the EEA. The upper and lower stripes represent the \texttt{Work1} and \texttt{Work2} registers, respectively. The symbols (h) and (l) indicate that the value $t'$ is split into its higher-order bits (h) and lower-order bits (l), which are placed at the corresponding positions on the two sides of the \texttt{Work2} register.}
\label{fig:regshare}
\end{figure}
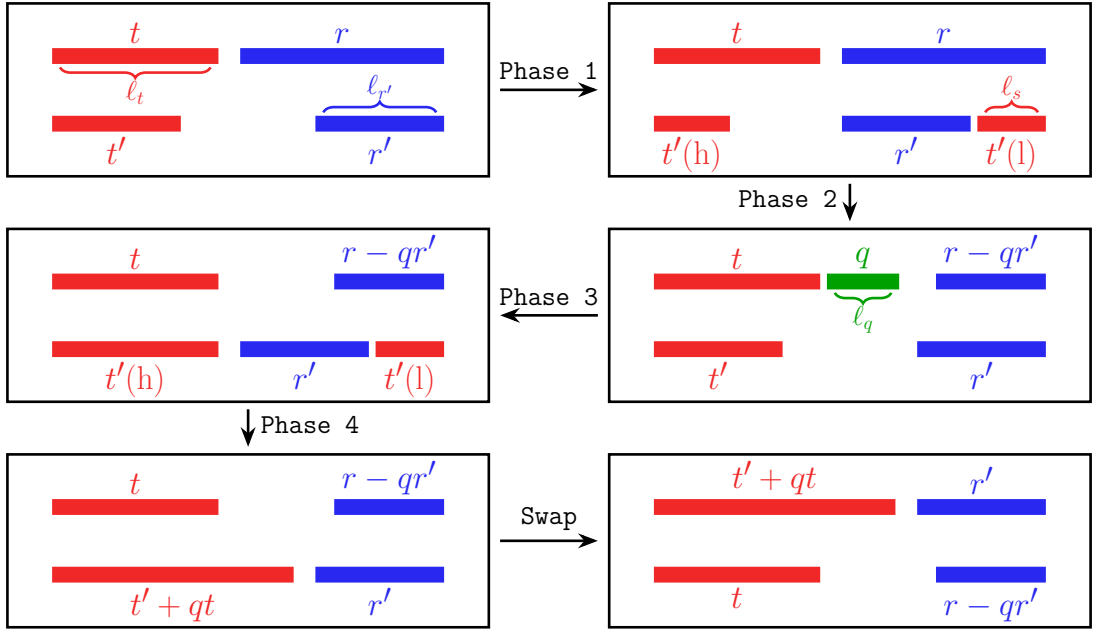

We extend this register-sharing idea further. Specifically, we observe that a single quantum register of $(n + 2)$ qubits can be used to store the triple $(r_{i-1}, t_i, q_i)$ simultaneously, including the intermediate states that arise during phases 2 and 3 of the algorithm. These phases are responsible for computing $q_i = \lfloor r_{i-1} / r_i \rfloor$ and updating the variables
\[
    r_{i-1} \leftarrow r_{i-1} - q_i r_i, t_{i-1} \leftarrow t_{i-1} + q_i t_i.
\]
The detailed allocation of quantum registers is described as follows, and illustrated in Figure \ref{fig:regshare}.
\begin{itemize}
    \item \textbf{\texttt{Work1} register (of $(n + 3)$ qubits).}  
    The left-most $\ell_t + 1$ qubits encode the value $t := t_i$ in little-endian order (here we append a zero to the rightmost position of $t$ in order to simplify the circuit implementation). The following $\ell_q$ qubits store the intermediate quotient value $q := q'$ (only the effective most significant bits) in phases 2 and 3, in big-endian order. The remaining qubits hold the value $r := r_{i-1}$ (or its updated form $r := r_{i-1} - q'r_i$) in big-endian order.

    \item \textbf{\texttt{Work2} register (of $(n + 3)$ qubits).}
    The right-most $\ell_{r'}$ qubits store $r' := r_i$ in big-endian order, while the remaining qubits store the value $t' := t_{i-1}$ (and its intermediate update $t' := t_{i-1} + q''t_i$) in little-endian order. During phases 2 and 3, this register may be circularly shifted left by $\ell_s$ positions. In such cases, the value $t'$ may span both ends of the register, effectively splitting into two contiguous parts.

    \item \textbf{\texttt{Length} registers.}  
    Four auxiliary registers are used to manage variable-length storage. Three of them each contain $\lfloor \log_2 n\rfloor + 2$ qubits and store the length indicators $\ell_t, \ell_q,$ and $\ell_{r'}$, respectively. The fourth register, of $\lfloor \log_2 n\rfloor + 3$ qubits, stores the shift counter $\ell_s$. These \texttt{Length} registers mark the logical boundaries within the \texttt{Work1} and \texttt{Work2} registers and enable coherent manipulation of superposed inputs of varying lengths.

    \item \textbf{\texttt{Control} registers.}  
    Two single-qubit \texttt{Phase} registers indicate the current phase of the four-phase algorithmic framework, and one single-qubit \texttt{Iter} register that records the parity of the current EEA iteration number. Additionally, there are two auxiliary single-qubit registers, \texttt{Sign} and \texttt{Ctrl}, used for intermediate flag and control operations.
\end{itemize}

We next describe the behavior of the system when performing comparison, addition, and subtraction operations on the register pairs $(r, r')$ and $(t, t')$. The arithmetic between $r$ and $r'$ is relatively straightforward: a left shift by $\ell_s$ positions corresponds to multiplying $r'$ by $2^{\ell_s}$. In contrast, the arithmetic involving $t$ and $t'$ requires a more careful interpretation. A convenient way to view this operation is that a left shift of $\ell_s$ positions can be regarded as dividing $t'$ by $2^{\ell_s}$, where the integer part is placed in the left-most portion of \texttt{Work2}, and the fractional part occupies the right-most portion. When $t$ is added to the integer part of $t'$ (which are properly aligned), the arithmetic relation can be expressed as $2^{-\ell_s}t' + t = 2^{-\ell_s}(t' + 2^{\ell_s}t)$.
Thus, the expression $t' + 2^{\ell_s}t$ represents the desired result, corresponding to a left shift by $\ell_s$ positions.

With this register allocation, one step of our space-efficient EEA proceeds as in Algorithm \ref{alg:regshare}. To illustrate how the proposed space-efficient EEA operates in practice, we also provide a concrete execution example for $p = 37$ and $x = 13$, shown in Table \ref{tab:run_example}.
\begin{algorithm}[htbp]
\small
\caption{One step of our EEA implementation with register sharing}
\begin{algorithmic}
\Require \texttt{Work1} register stores $\ket{t, q, r}$, \texttt{Work2} register stores $\ket{t', r'}$
\Require \texttt{Length} registers store $\ket{\ell_t, \ell_q, \ell_{r'}, \ell_s}$
\Require \texttt{Control} registers store $\ket{\texttt{Phase1}, \texttt{Phase2}, \texttt{Iter}, \texttt{Sign}}$.

\If{$(\texttt{Phase1}, \texttt{Phase2}) = (0, 0)$} \Comment{Arithmetic logic for the four phases}
    \State Perform a one-position left shift on \texttt{Work2}
    \State $\ell_s \gets \ell_s + 1$
    \State $\texttt{Sign} \gets \texttt{Sign} \oplus (r < 2^{\ell_s}r')$ \Comment{Act on qubits $(\ell_t + \ell_q + 2)$ through $(n + 3 - \ell_s)$}
\ElsIf{$(\texttt{Phase1}, \texttt{Phase2}) = (0, 1)$}
    \State Perform a one-position right shift on \texttt{Work2}
    \State $\ell_s \gets \ell_s - 1$, $\ell_q \gets \ell_q + 1$
    \State $(\texttt{Sign}, r) \gets (\texttt{Sign}, r) - 2^{\ell_s}r'$ \Comment{Use register \texttt{Sign} to store the sign of the subtraction result}
    \If{$\texttt{Sign} = 1$}
        \State $r \gets r + 2^{\ell_s}r'$ \Comment{Ignore overflows}
    \EndIf
    \State $\texttt{Sign} \gets \texttt{Sign} \oplus 1$
    \State Swap $\texttt{Sign}$ with the $(\ell_t + \ell_q + 1)$-th qubit of \texttt{Work1} \Comment{The least significant work qubit of $q$}
\ElsIf{$(\texttt{Phase1}, \texttt{Phase2}) = (1, 0)$}
    \State Swap $\texttt{Sign}$ with the $(\ell_t + \ell_q + 1)$-th qubit of \texttt{Work1}
    \If{$\texttt{Sign} = 0$}
        \State $t' \gets t' - 2^{\ell_s}t$ \Comment{Act on left-most $(\ell_t + 1)$ qubits}
    \EndIf
    \State $\texttt{Sign} \gets \texttt{Sign} \oplus 1$
    \State $(\texttt{Sign}, t') \gets (\texttt{Sign}, t') + 2^{\ell_s}t$ \Comment{Use register \texttt{Sign} to store the sign of the addition result}
    \State Perform a one-position left shift on \texttt{Work2}
    \State $\ell_s \gets \ell_s + 1$, $\ell_q \gets \ell_q - 1$
\ElsIf{$(\texttt{Phase1}, \texttt{Phase2}) = (1, 1)$}
    \State $\texttt{Sign} \gets \texttt{Sign} \oplus (t' \ge 2^{\ell_s}t)$ \Comment{Same qubit range as above}
    \State Perform a one-position right shift on \texttt{Work2}
    \State $\ell_s \gets \ell_s - 1$
\EndIf

\If{$\ell_q = 0$ \textbf{and} $\ell_{r'} > 0$} \Comment{Phase update logic; $\ell_{r'} = 0$ indicates algorithm termination}
    \State $\texttt{Phase2} \gets \texttt{Phase2} \oplus \texttt{Sign} \oplus \texttt{Phase1}$, $\texttt{Sign} \gets \texttt{Sign} \oplus \texttt{Phase2}$
\EndIf
\If{$\ell_s = 0$}
    \State $\texttt{Phase1} \gets \texttt{Phase1} \oplus 1$, $\texttt{Phase2} \gets \texttt{Phase2} \oplus 1$
\EndIf

\If{$\ell_q = 0$ \textbf{and} $\ell_s = 0$} \Comment{Register swapping at the end of one EEA iteration}
    \State Swap \texttt{Work1} and \texttt{Work2}
    \State Update $\ell_t$ to the bit length of the new $t$ \Comment{Act on the left-most $(n + 3 - \ell_{r'})$ qubits of \texttt{Work1} and \texttt{Work2}}
    \State Update $\ell_{r'}$ to the bit length of the new $r'$ \Comment{Act on the right-most $(n + 2 - \ell_t)$ qubits of \texttt{Work1} and \texttt{Work2}}
    \State $\texttt{Iter} \gets \texttt{Iter} \oplus 1$
\EndIf
\end{algorithmic}
\label{alg:regshare}
\end{algorithm}

\begin{table}[htbp]
\centering
\footnotesize
\begin{tabular}{c c c c c c c c c c c c c c c c c}
\hline
\textbf{Step} & \texttt{Work1} & \texttt{Work2} & $t$ & $q$ & $r$ & $t'$ & $r'$ & $\ell_t$ & $\ell_q$ & $\ell_{r'}$ & $\ell_s$ & \texttt{Phase1} & \texttt{Phase2} & \texttt{Iter} & \texttt{Sign} \\
\hline
0 & \verb+10||0100101+ & \verb+00000|1101|+ & 1 & 0 & 37 & 0 & 13 & 1 & 0 & 4 & 0 & 0 & 0 & 0 & 0 \\
\hline
1 & \verb+10||0100101+ & \verb+0000|1101|0+ & 1 & 0 & 37 & 0 & 13 & 1 & 0 & 4 & 1 & 0 & 0 & 0 & 0 \\
\hline
2 & \verb+10||0100101+ & \verb+000|1101|00+ & 1 & 0 & 37 & 0 & 13 & 1 & 0 & 4 & 2 & 0 & 1 & 0 & 0 \\
\hline
3 & \verb+10|1|001011+ & \verb+0000|1101|0+ & 1 & 2 & 11 & 0 & 13 & 1 & 1 & 4 & 1 & 0 & 1 & 0 & 0 \\
\hline
4 & \verb+10|10|01011+ & \verb+00000|1101|+ & 1 & 2 & 11 & 0 & 13 & 1 & 2 & 4 & 0 & 1 & 0 & 0 & 0 \\
\hline
5 & \verb+10|1|001011+ & \verb+0000|1101|0+ & 1 & 2 & 11 & 0 & 13 & 1 & 1 & 4 & 1 & 1 & 0 & 0 & 0 \\
\hline
6 & \verb+10||0001011+ & \verb+000|1101|01+ & 1 & 0 & 11 & 2 & 13 & 1 & 0 & 4 & 2 & 1 & 1 & 0 & 1 \\
\hline
7 & \verb+10||0001011+ & \verb+1000|1101|0+ & 1 & 0 & 11 & 2 & 13 & 1 & 0 & 4 & 1 & 1 & 1 & 0 & 0 \\
\hline
8 & \verb+010||001101+ & \verb+10000|1011|+ & 2 & 0 & 13 & 1 & 11 & 2 & 0 & 4 & 0 & 0 & 0 & 1 & 0 \\
\hline
9 & \verb+010||001101+ & \verb+0000|1011|1+ & 2 & 0 & 13 & 1 & 11 & 2 & 0 & 4 & 1 & 0 & 1 & 1 & 0 \\
\hline
10 & \verb+010|1|00010+ & \verb+10000|1011|+ & 2 & 1 & 2 & 1 & 11 & 2 & 1 & 4 & 0 & 1 & 0 & 1 & 0 \\
\hline
11 & \verb+010||000010+ & \verb+1000|1011|1+ & 2 & 0 & 2 & 3 & 11 & 2 & 0 & 4 & 1 & 1 & 1 & 1 & 1 \\
\hline
12 & \verb+110||001011+ & \verb+0100000|10|+ & 3 & 0 & 11 & 2 & 2 & 2 & 0 & 2 & 0 & 0 & 0 & 0 & 0 \\
\hline
13 & \verb+110||001011+ & \verb+100000|10|0+ & 3 & 0 & 11 & 2 & 2 & 2 & 0 & 2 & 1 & 0 & 0 & 0 & 0 \\
\hline
14 & \verb+110||001011+ & \verb+00000|10|01+ & 3 & 0 & 11 & 2 & 2 & 2 & 0 & 2 & 2 & 0 & 0 & 0 & 0 \\
\hline
15 & \verb+110||001011+ & \verb+0000|10|010+ & 3 & 0 & 11 & 2 & 2 & 2 & 0 & 2 & 3 & 0 & 1 & 0 & 0 \\
\hline
16 & \verb+110|1|00011+ & \verb+00000|10|01+ & 3 & 4 & 3 & 2 & 2 & 2 & 1 & 2 & 2 & 0 & 1 & 0 & 0 \\
\hline
17 & \verb+110|10|0011+ & \verb+100000|10|0+ & 3 & 4 & 3 & 2 & 2 & 2 & 2 & 2 & 1 & 0 & 1 & 0 & 0 \\
\hline
18 & \verb+110|101|001+ & \verb+0100000|10|+ & 3 & 5 & 1 & 2 & 2 & 2 & 3 & 2 & 0 & 1 & 0 & 0 & 0 \\
\hline
19 & \verb+110|10|0001+ & \verb+010000|10|1+ & 3 & 4 & 1 & 5 & 2 & 2 & 2 & 2 & 1 & 1 & 0 & 0 & 0 \\
\hline
20 & \verb+110|1|00001+ & \verb+10000|10|10+ & 3 & 4 & 1 & 5 & 2 & 2 & 1 & 2 & 2 & 1 & 0 & 0 & 0 \\
\hline
21 & \verb+110||000001+ & \verb+0100|10|100+ & 3 & 0 & 1 & 17 & 2 & 2 & 0 & 2 & 3 & 1 & 1 & 0 & 1 \\
\hline
22 & \verb+110||000001+ & \verb+00100|10|10+ & 3 & 0 & 1 & 17 & 2 & 2 & 0 & 2 & 2 & 1 & 1 & 0 & 0 \\
\hline
23 & \verb+110||000001+ & \verb+000100|10|1+ & 3 & 0 & 1 & 17 & 2 & 2 & 0 & 2 & 1 & 1 & 1 & 0 & 0 \\
\hline
24 & \verb+100010||010+ & \verb+11000000|1|+ & 17 & 0 & 2 & 3 & 1 & 5 & 0 & 1 & 0 & 0 & 0 & 1 & 0 \\
\hline
25 & \verb+100010||010+ & \verb+1000000|1|1+ & 17 & 0 & 2 & 3 & 1 & 5 & 0 & 1 & 1 & 0 & 0 & 1 & 0 \\
\hline
26 & \verb+100010||010+ & \verb+000000|1|11+ & 17 & 0 & 2 & 3 & 1 & 5 & 0 & 1 & 2 & 0 & 1 & 1 & 0 \\
\hline
27 & \verb+100010|1|00+ & \verb+1000000|1|1+ & 17 & 2 & 0 & 3 & 1 & 5 & 1 & 1 & 1 & 0 & 1 & 1 & 0 \\
\hline
28 & \verb+100010|10|0+ & \verb+11000000|1|+ & 17 & 2 & 0 & 3 & 1 & 5 & 2 & 1 & 0 & 1 & 0 & 1 & 0 \\
\hline
29 & \verb+100010|1|00+ & \verb+1000000|1|1+ & 17 & 2 & 0 & 3 & 1 & 5 & 1 & 1 & 1 & 1 & 0 & 1 & 0 \\
\hline
30 & \verb+100010||000+ & \verb+100100|1|10+ & 17 & 0 & 0 & 37 & 1 & 5 & 0 & 1 & 2 & 1 & 1 & 1 & 1 \\
\hline
31 & \verb+100010||000+ & \verb+0100100|1|1+ & 17 & 0 & 0 & 37 & 1 & 5 & 0 & 1 & 1 & 1 & 1 & 1 & 0 \\
\hline
32 & \verb+Terminated+ & \verb+Terminated+ & 37 & 0 & 1 & 17 & 0 & 6 & 0 & 0 & 0 & 0 & 0 & 0 & 0 \\
\hline
33 & \verb+Terminated+ & \verb+Terminated+ & 37 & 0 & 1 & 17 & 0 & 6 & 0 & 0 & 1 & 0 & 0 & 0 & 0 \\
\hline
34 & \verb+Terminated+ & \verb+Terminated+ & 37 & 0 & 1 & 17 & 0 & 6 & 0 & 0 & 2 & 0 & 0 & 0 & 0 \\
\hline
35 & \verb+Terminated+ & \verb+Terminated+ & 37 & 0 & 1 & 17 & 0 & 6 & 0 & 0 & 3 & 0 & 0 & 0 & 0 \\
\hline
36 & \verb+Terminated+ & \verb+Terminated+ & 37 & 0 & 1 & 17 & 0 & 6 & 0 & 0 & 4 & 0 & 0 & 0 & 0 \\
\hline
\end{tabular}
\caption{An execution example of our space-efficient EEA for $p = 37$ and $x = 13$.
The table mainly illustrates how the two $(n+3)$-qubit registers \texttt{Work1} and \texttt{Work2} are allocated during the execution to store the variables $t, q, r$ and $t', r'$, respectively.}
\label{tab:run_example}
\end{table}

For each step, Table \ref{tab:run_example} shows how the contents of the two \texttt{Work} registers evolve and how different variables are mapped onto \texttt{Work1} and \texttt{Work2} under the register-sharing strategy.
In particular, the variable $t'$ stored in \texttt{Work2} is divided into two parts that occupy the left-most and right-most portions of the register. The division symbols are not explicitly implemented in the quantum circuit; they are all illustrative and inferred from the corresponding \texttt{Length} registers, used to visualize the register allocation.
In the actual quantum implementation, all registers are in superposition, and the location of these division symbols may vary across different input values.

%% file: sec_EEA/sub_opt.tex
\subsection{Algorithmic optimizations}
\label{subsec:opt}

The baseline implementation of one step, as described in the previous subsection, was constructed to be fully reversible, ensuring that every state transformation can be realized as a unitary operation on a quantum computer. However, this direct implementation contains multiple redundant operations that unnecessarily increase both the circuit depth and gate count. In this subsection, we present a series of algorithmic optimizations applied to the baseline design in order to reduce the quantum circuit depth and overall gate complexity.

\paragraph{Removing redundant arithmetic and shift operations.}
In the baseline version, several arithmetic and shift operations between different phases are functionally redundant. Specifically, we have:
\[
\begin{split}
    \text{In phases 1 and 2:} \quad & \text{comparison, addition, subtraction between } (r, 2^{\ell_s}r'); \\
    \text{In phases 3 and 4:} \quad & \text{comparison, addition, subtraction between } (t', 2^{\ell_s}t); \\
    \text{In phases 1 and 3:} \quad & \text{a one-position left shift on } \texttt{Work2}; \\
    \text{In phases 2 and 4:} \quad & \text{a one-position right shift on } \texttt{Work2}. \\
\end{split}
\]
By combining the control conditions of these operations (or equivalently, the corresponding quantum control wires), we can eliminate redundant operations from the circuit. This optimization results in a total of two location-controlled addition and subtraction, and two location-controlled swaps in one step. Here, \emph{location-control} refers to arithmetic operations applied conditionally on specific positions of the two \texttt{Work} registers, with the active positions indicated by the \texttt{Length} registers.

\paragraph{Merging location-controlled swaps.}
To further decrease the circuit depth and gate count, we observe that a single location-controlled swap requires $O(n\log_2 n)$ quantum gates. When two such swaps are merged into a single operation, two additional position shifts on \texttt{Work2} are inevitably introduced. Fortunately, this trade-off is beneficial: each position shift requires only $O(n)$ gates, making the merging optimization asymptotically advantageous.

\paragraph{Algorithm pseudocode and overall complexity.}
After applying the above optimizations, the overall asymptotic quantum resource estimation per step is summarized as follows:
\begin{itemize}
    \item Five location-controlled arithmetic operations, including one swap, two additions, and two subtraction; along with four length updates at the termination step of an EEA iteration. Each of these operations costs $O(n\log_2 n)$ gates in a straightforward implementation.
    \item Four position shifts and one $(n + 3)$-qubit SWAP, each requiring $O(n)$ quantum gates.
\end{itemize}

%\tyl{Prove a non-asymptotic bound here?}

With these optimizations, one step of our space-efficient EEA proceeds as in Algorithm \ref{alg:opt}.
\begin{algorithm}[htbp]
\small
\caption{Optimized one step of our EEA implementation}
\begin{algorithmic}
\Require \texttt{Work1} register stores $\ket{t, q, r}$, \texttt{Work2} register stores $\ket{t', r'}$;
\Require \texttt{Length} registers store $\ket{\ell_t, \ell_q, \ell_{r'}, \ell_s}$;
\Require \texttt{Control} registers store $\ket{\texttt{Phase1}, \texttt{Phase2}, \texttt{Iter}, \texttt{Sign}}$.

\If{$\texttt{Phase1} = 0$} \Comment{Pre-shift operations}
    \State Perform a one-position left shift on \texttt{Work2}, and update $\ell_s \gets \ell_s + 1$
    \If{$\texttt{Phase2} = 1$}
        \State Perform a two-position right shift on \texttt{Work2}, and update $\ell_s \gets \ell_s - 2$
    \EndIf
    
    \State $(\texttt{Sign}, r) \gets (\texttt{Sign}, r) - 2^{\ell_s}r'$ \Comment{Arithmetic block 1: location-controlled subtraction on $r$'s}
    \If{$\texttt{Phase2} = 1$}
        \State $\texttt{Sign} \gets \texttt{Sign} \oplus 1$
    \EndIf
    \If{$\texttt{Phase2} = 0$ \textbf{or} $\texttt{Sign} = 0$}
        \State $r \gets r + 2^{\ell_s}r'$
    \EndIf
\EndIf

\If{$\texttt{Phase1} \oplus \texttt{Phase2} = 1$} \Comment{Arithmetic block 2: location-controlled swap}
    \State Swap $\texttt{Sign}$ with the $(\ell_t + \ell_q + 1)$-th qubit of \texttt{Work1}
    \If{$\texttt{Phase1} = 1$}
        \State $\ell_q \gets \ell_q - 1$
    \Else
        \State $\ell_q \gets \ell_q + 1$
    \EndIf
\EndIf

\If{$\texttt{Phase1} = 1$} \Comment{Arithmetic block 3: location-controlled addition on $t$'s}
    \If{$\texttt{Phase2} = 1$ \textbf{or} $\texttt{Sign} = 0$}
        \State $t' \gets t' - 2^{\ell_s}t$
    \EndIf
    \State $\texttt{Sign} \gets \texttt{Sign} \oplus 1$
    \State $(\texttt{Sign}, t') \gets (\texttt{Sign}, t') + 2^{\ell_s}t$

    \State Perform a one-position left shift on \texttt{Work2}, and update $\ell_s \gets \ell_s + 1$ \Comment{Post-shift operations}
    \If{$\texttt{Phase2} = 1$}
        \State Perform a two-position right shift on \texttt{Work2}, and update $\ell_s \gets \ell_s - 2$
    \EndIf
\EndIf

\If{$\ell_q = 0$ \textbf{and} $\ell_{r'} > 0$} \Comment{Phase update logic; $\ell_{r'} = 0$ indicates algorithm termination}
    \State $\texttt{Phase2} \gets \texttt{Phase2} \oplus \texttt{Sign} \oplus \texttt{Phase1}$, $\texttt{Sign} \gets \texttt{Sign} \oplus \texttt{Phase2}$
\EndIf
\If{$\ell_s = 0$}
    \State $\texttt{Phase1} \gets \texttt{Phase1} \oplus 1$, $\texttt{Phase2} \gets \texttt{Phase2} \oplus 1$
\EndIf

\If{$\ell_q = 0$ \textbf{and} $\ell_s = 0$} \Comment{Register swapping at the end of one EEA iteration}
    \State Swap \texttt{Work1} and \texttt{Work2}
    \State Update $\ell_t$ to the bit length of the new $t$, $\ell_{r'}$ to the bit length of the new $r'$
    \State $\texttt{Iter} \gets \texttt{Iter} \oplus 1$
\EndIf
\end{algorithmic}
\label{alg:opt}
\end{algorithm}

%% file: sec_circuit/sec_circuit.tex
\section{Quantum Circuit Implementation of Modular Inversion}
\label{sec:circuit}

In this section, we present the explicit quantum circuit implementation of a single step of the space-efficient EEA with register sharing, as described in Algorithm \ref{alg:opt}.
Throughout the section, we use the same quantum register notation as in the algorithm.

The full circuit construction is, however, too complex to be depicted within a small number of circuit diagrams.
This difficulty is mainly due to the presence of location-controlled arithmetic operations, which are highly non-standard.
To present the construction in a relatively more explicit manner, we first describe the overall circuit structure in Section \ref{subsec:all_circuit}, where the location-controlled arithmetic operations are treated as circuit blocks; we then provide a detailed implementation of these operations in Section \ref{subsec:blocks}. In all circuit diagrams, the black triangle denotes the output wire that carries the result of the corresponding circuit block; and we denote $\ell = \lfloor \log_2 n \rfloor$ for convenience.

\input{sec_circuit/sub_all}
\input{sec_circuit/sub_block}

\input{sec_circuit/sub_active}

%% file: sec_circuit/sub_all.tex
\subsection{Overall circuit implementation of one step}
\label{subsec:all_circuit}

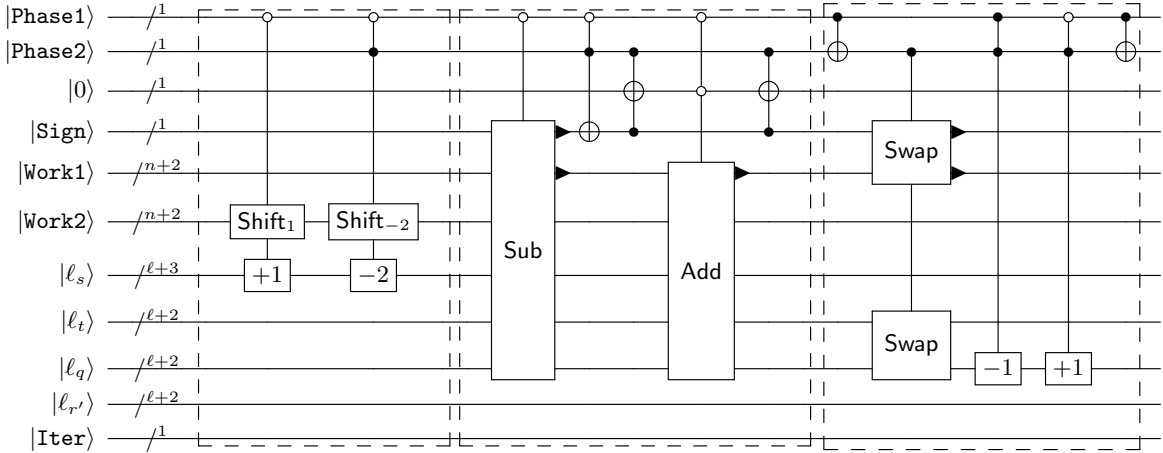
\begin{figure}[ht]
\footnotesize
\centering
\(\Qcircuit @C=1.0em @R=0.8em {
    & \lstick{\ket{\texttt{Phase1}}} & \qw & \qw\slash^1 & \qw & \qw 
    & \ctrlo{5} & \ctrlo{1} & \qw & \qw & \ctrlo{3} 
    & \ctrlo{1} & \qw & \ctrlo{2} & \qw & \qw 
    & \ctrl{1} & \qw & \ctrl{1} & \ctrlo{1} & \ctrl{1} 
    & \qw \\
    & \lstick{\ket{\texttt{Phase2}}} & \qw & \qw\slash^1 & \qw & \qw 
    & \qw & \ctrl{4} & \qw & \qw & \qw 
    & \ctrl{2} & \ctrl{1} & \qw & \ctrl{1} & \qw 
    & \targ & \ctrl{2} & \ctrl{7} & \ctrl{7} & \targ 
    & \qw \\
    & \lstick{\ket{0}} & \qw & \qw\slash^1 & \qw & \qw 
    & \qw & \qw & \qw & \qw & \qw 
    & \qw & \targ & \ctrlo{2} & \targ & \qw 
    & \qw & \qw & \qw & \qw & \qw 
    & \qw \\
    & \lstick{\ket{\texttt{Sign}}} & \qw & \qw\slash^1 & \qw & \qw 
    & \qw & \qw & \qw & \qw & \multigateouttwohi{5}{\mathsf{Sub}} 
    & \targ & \ctrl{-1} & \qw & \ctrl{-1} & \qw 
    & \qw & \multigateouttwo{1}{\mathsf{Swap}} & \qw & \qw & \qw 
    & \qw \\
    & \lstick{\ket{\texttt{Work1}}} & \qw & \qw\slash^{n+2} & \qw & \qw 
    & \qw & \qw & \qw & \qw & \ghost{\mathsf{Sub}} 
    & \qw & \qw & \multigateout{4}{\mathsf{Add}} & \qw & \qw 
    & \qw & \ghost{\mathsf{Swap}}\qwx[3] & \qw & \qw & \qw 
    & \qw \\
    & \lstick{\ket{\texttt{Work2}}} & \qw & \qw\slash^{n+2} & \qw & \qw 
    & \gate{\mathsf{Shift}_1}\qwx[1] & \gate{\mathsf{Shift}_{-2}}\qwx[1] & \qw & \qw & \ghost{\mathsf{Sub}} 
    & \qw & \qw & \ghost{\mathsf{Add}} & \qw & \qw
    & \qw & \qw & \qw & \qw & \qw 
    & \qw \\
    & \lstick{\ket{\ell_s}} & \qw & \qw\slash^{\ell + 3} & \qw & \qw 
    & \gate{+1} & \gate{-2} & \qw & \qw & \ghost{\mathsf{Sub}} 
    & \qw & \qw & \ghost{\mathsf{Add}} & \qw & \qw
    & \qw & \qw & \qw & \qw & \qw 
    & \qw \\
    & \lstick{\ket{\ell_t}} & \qw & \qw\slash^{\ell + 2} & \qw & \qw 
    & \qw & \qw & \qw & \qw & \ghost{\mathsf{Sub}} 
    & \qw & \qw & \ghost{\mathsf{Add}} & \qw & \qw 
    & \qw & \multigate{1}{\mathsf{Swap}} & \qw & \qw & \qw 
    & \qw \\
    & \lstick{\ket{\ell_q}} & \qw & \qw\slash^{\ell + 2} & \qw & \qw 
    & \qw & \qw & \qw & \qw & \ghost{\mathsf{Sub}} 
    & \qw & \qw & \ghost{\mathsf{Add}} & \qw & \qw 
    & \qw & \ghost{\mathsf{Swap}} & \gate{-1} & \gate{+1} & \qw 
    & \qw \\
    & \lstick{\ket{\ell_{r'}}} & \qw & \qw\slash^{\ell + 2} & \qw & \qw 
    & \qw & \qw & \qw & \qw & \qw 
    & \qw & \qw & \qw & \qw & \qw 
    & \qw & \qw & \qw & \qw & \qw 
    & \qw \\
    \push{\rule{0em}{1.2em}} & \lstick{\ket{\texttt{Iter}}} & \qw & \qw\slash^{1} & \qw & \qw 
    & \qw & \qw & \qw & \qw & \qw 
    & \qw & \qw & \qw & \qw & \qw 
    & \qw & \qw & \qw & \qw & \qw 
    & \qw
    \gategroup{1}{6}{11}{9}{0.6em}{--}
    \gategroup{1}{10}{11}{16}{0.6em}{--}
    \gategroup{1}{17}{11}{21}{0.9em}{--}
}\)
\caption{Overall circuit implementation (Part 1) of a single iteration of our space-efficient EEA.
The three dashed boxes, from left to right, represent:
(1) pre-shift operations;
(2) location-controlled subtraction on $r$'s;
(3) location-controlled swap.}
\label{fig:all_circuit_1}
\end{figure}

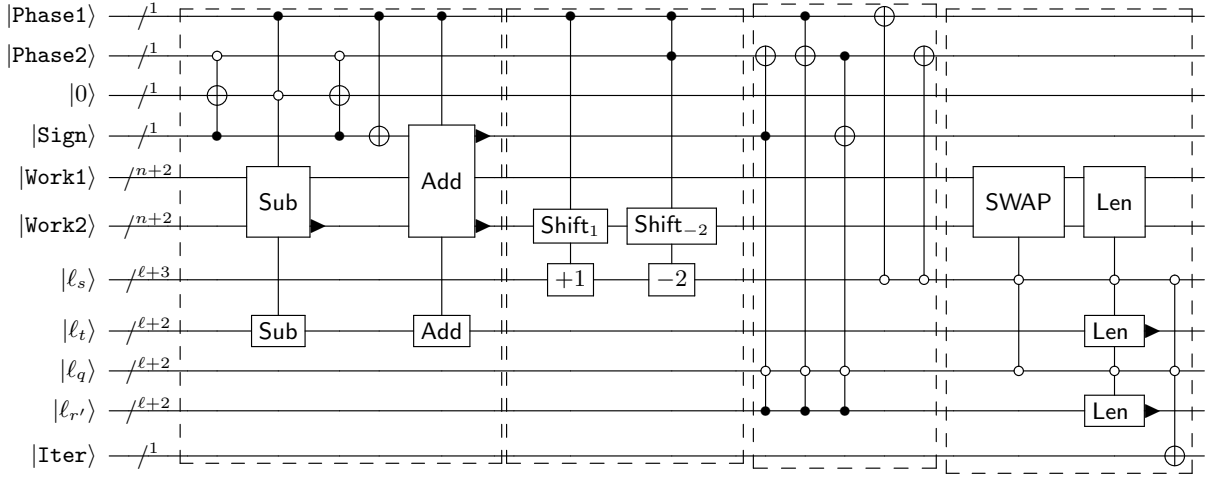
\begin{figure}[ht]
\footnotesize
\centering
\(\Qcircuit @C=0.8em @R=0.8em {
    & \lstick{\ket{\texttt{Phase1}}} & \qw & \qw\slash^1 & \qw & \qw 
    & \qw & \ctrl{2} & \qw & \ctrl{3} & \ctrl{3} 
    & \qw & \qw & \ctrl{5} & \ctrl{1} & \qw 
    & \qw & \ctrl{1} & \qw & \targ & \qw
    & \qw & \qw & \qw & \qw & \qw \\
    & \lstick{\ket{\texttt{Phase2}}} & \qw & \qw\slash^1 & \qw & \qw 
    & \ctrlo{1} & \qw & \ctrlo{1} & \qw & \qw 
    & \qw & \qw & \qw & \ctrl{4} & \qw 
    & \targ & \targ & \ctrl{2} & \qw & \targ 
    & \qw & \qw & \qw & \qw & \qw \\
    & \lstick{\ket{0}} & \qw & \qw\slash^1 & \qw & \qw 
    & \targ & \ctrlo{2} & \targ & \qw & \qw 
    & \qw & \qw & \qw & \qw & \qw 
    & \qw & \qw & \qw & \qw & \qw 
    & \qw & \qw & \qw & \qw & \qw \\
    & \lstick{\ket{\texttt{Sign}}} & \qw & \qw\slash^1 & \qw & \qw 
    & \ctrl{-1} & \qw & \ctrl{-1} & \targ & \multigateouttwo{2}{\mathsf{Add}} 
    & \qw & \qw & \qw & \qw & \qw 
    & \ctrl{-2} & \qw & \targ & \qw & \qw 
    & \qw & \qw & \qw & \qw & \qw \\
    & \lstick{\ket{\texttt{Work1}}} & \qw & \qw\slash^{n+2} & \qw & \qw 
    & \qw & \ghost{\mathsf{Sub}} & \qw & \qw & \ghost{\mathsf{Add}} 
    & \qw & \qw & \qw & \qw & \qw 
    & \qw & \qw & \qw & \qw & \qw 
    & \qw & \multigate{1}{\mathsf{SWAP}} & \multigate{1}{\mathsf{Len}} & \qw & \qw \\
    & \lstick{\ket{\texttt{Work2}}} & \qw & \qw\slash^{n+2} & \qw & \qw 
    & \qw & \multigateout{-1}{\mathsf{Sub}}\qwx[2] & \qw & \qw & \ghost{\mathsf{Add}}\qwx[2] 
    & \qw & \qw & \gate{\mathsf{Shift}_1}\qwx[1] & \gate{\mathsf{Shift}_{-2}}\qwx[1] & \qw 
    & \qw & \qw & \qw & \qw & \qw 
    & \qw & \ghost{\mathsf{SWAP}} & \ghost{\mathsf{Len}} & \qw & \qw \\
    & \lstick{\ket{\ell_s}} & \qw & \qw\slash^{\ell + 3} & \qw & \qw 
    & \qw & \qw & \qw & \qw & \qw 
    & \qw & \qw & \gate{+1} & \gate{-2} & \qw 
    & \qw & \qw & \qw & \ctrlo{-6} & \ctrlo{-5} 
    & \qw & \ctrlo{-1} & \ctrlo{-1} & \ctrlo{2} & \qw \\
    & \lstick{\ket{\ell_t}} & \qw & \qw\slash^{\ell + 2} & \qw & \qw 
    & \qw & \gate{\mathsf{Sub}} & \qw & \qw & \gate{\mathsf{Add}} 
    & \qw & \qw & \qw & \qw & \qw 
    & \qw & \qw & \qw & \qw & \qw 
    & \qw & \qw & \gateout{\mathsf{Len}}\qwx[-1] & \qw & \qw \\
    & \lstick{\ket{\ell_q}} & \qw & \qw\slash^{\ell + 2} & \qw & \qw 
    & \qw & \qw & \qw & \qw & \qw 
    & \qw & \qw & \qw & \qw & \qw 
    & \ctrlo{-5} & \ctrlo{-7} & \ctrlo{-5} & \qw & \qw 
    & \qw & \ctrlo{-2} & \ctrlo{-1} & \ctrlo{2} & \qw \\
    & \lstick{\ket{\ell_{r'}}} & \qw & \qw\slash^{\ell + 2} & \qw & \qw 
    & \qw & \qw & \qw & \qw & \qw 
    & \qw & \qw & \qw & \qw & \qw 
    & \ctrl{-1} & \ctrl{-1} & \ctrl{-1} & \qw & \qw 
    & \qw & \qw & \gateout{\mathsf{Len}}\qwx[-1] & \qw & \qw \\
    & \lstick{\ket{\texttt{Iter}}} & \qw & \qw\slash^1 & \qw & \qw 
    & \qw & \qw & \qw & \qw & \qw 
    & \qw & \qw & \qw & \qw & \qw 
    & \qw & \qw & \qw & \qw & \qw 
    & \qw & \qw & \qw & \targ & \qw
    \gategroup{1}{6}{11}{12}{0.6em}{--}
    \gategroup{1}{13}{11}{16}{0.6em}{--}
    \gategroup{1}{17}{11}{21}{1.0em}{--}
    \gategroup{1}{22}{11}{25}{0.6em}{--}
}\)
\caption{Overall circuit implementation (Part 2) of a single iteration of our space-efficient EEA.
The four dashed boxes, from left to right, represent:
(1) location-controlled addition on $t$'s;
(2) post-shift operations;
(3) phase update;
(4) swapping \texttt{Work1} and \texttt{Work2}, together with the corresponding length updates at the end of one EEA iteration.}
\label{fig:all_circuit_2}
\end{figure}

According to Algorithm \ref{alg:opt}, the overall circuit implementation of a single iteration of our space-efficient EEA is shown in Figures \ref{fig:all_circuit_1} and \ref{fig:all_circuit_2}, presented in a continuous layout.
Both circuit diagrams are divided into several algorithmic components, indicated by dashed boxes.
Each component corresponds exactly to one arithmetic block specified in Algorithm \ref{alg:opt}.

In both circuit diagrams, $\mathsf{Shift}$ denotes a position shift of the workspace: a positive value represents a left shift, while a negative value represents a right shift.
The circuit blocks $\mathsf{Add}$, $\mathsf{Sub}$, and $\mathsf{Swap}$ denote location-controlled addition, subtraction, and swap operations, respectively.
At the end of each EEA iteration, the circuit block labeled $\mathsf{SWAP}$ represents a full SWAP operation applied to all qubits in the registers \texttt{Work1} and \texttt{Work2}; this block is distinct from the $\mathsf{Swap}$ block to avoid ambiguity.
Finally, the $\mathsf{Len}$ block denotes the update operations on the length registers.

%% file: sec_circuit/sub_block.tex
\subsection{Location-controlled operations}
\label{subsec:blocks}

In this subsection, the two \texttt{Work} registers are indexed, from left to right, by $u_1, u_2, \cdots, u_{n+3}$ and $v_1, v_2, \ldots, v_{n+3}$, respectively.
In practice, we recommend storing the truth value minus one in all \texttt{Length} registers. 
The reason is that Algorithm \ref{alg:opt} contains controlled operations that are performed when the value stored in the \texttt{Length} register equals zero. 
By encoding the value as the truth value minus one, we can detect whether the original truth value equals zero by looking at the sign bit of the \texttt{Length} register, avoiding the need to implement a multi-qubit controlled operation over $\log_2 n$ qubits.

\begin{figure}[ht]
\centering
\footnotesize 
\(\Qcircuit @C=1.2em @R=1.4em {
    & \lstick{\ket{u_k}} & \qw & \qw & \qw & \qw 
    & \qw & \qw & \qw & \qw & \qw 
    & \cdots & & \qw & \qw & \qw
    & \qswap & \qw & \qw & \qw \\
    & \lstick{\ket{u_{k+1}}} & \qw & \qw & \qw & \qw 
    & \qw & \qw & \qw & \qw & \qw 
    & \cdots & & \qswap & \qw & \qswap
    & \qw & \qw & \qw & \qw \\
    & \lstick{\vdots\ } & & & & & & & & &
    & \vdots & & & & & & & & \\
    & & & & & & & & & & & & & & & & & & \\
    & \lstick{\ket{u_{K-1}}} & \qw & \qw & \qw & \qw 
    & \qw & \qswap & \qw & \qswap & \qw 
    & \cdots & & \qw & \qw & \qw
    & \qw & \qw & \qw & \qw \\
    & \lstick{\ket{u_{K}}} & \qw & \qw & \qw & \qw 
    & \qswap & \qw & \qw & \qw & \qw 
    & \cdots & & \qw & \qw & \qw
    & \qw & \qw & \qw & \qw \\
    & \lstick{\ket{\ell_t-1}} & \qw\slash^{\ell+2} & \qw & \multigateouttwolow{2}{+} & \qw 
    & \qw & \qw & \qw & \qw & \qw
    & \cdots & & \qw & \qw & \qw
    & \qw & \qw & \multigateouttwolow{2}{-} & \qw \\
    & \lstick{\ket{\mathrm{sign}_{\ell_q}}} & \qw & \qw & \ghost{+}
    & \multigate{1}{-K^*}
    & \ctrlo{-2} & \ctrl{-3} & \multigate{1}{+1} & \ctrl{-3} & \qw
    & \cdots & & \ctrl{-6} & \multigate{1}{+1} & \ctrl{-6} 
    & \ctrl{-7} & \multigate{1}{+k^*} & \ghost{-} & \qw \\
    & \lstick{\ket{\ell_q - 1}} & \qw\slash^{\ell+1} & \qw & \ghost{+}
    & \ghost{-K^*} 
    & \qw & \qw & \ghost{+1} & \qw & \qw 
    & \cdots & & \qw & \ghost{+1} & \qw
    & \qw & \ghost{+k^*} & \ghost{-} & \qw \\
    & \lstick{\ket{\texttt{Sign}}} & \qw & \qw & \qw & \qw
    & \qswap & \qswap & \qw & \qswap & \qw
    & \cdots & & \qswap & \qw & \qswap 
    & \qswap & \qw & \qw & \qw \\
    & \lstick{\ket{\mathrm{\texttt{Ctrl}}}} & \qw & \qw & \qw & \qw
    & \ctrl{-3} & \ctrl{-3} & \qw & \ctrl{-3} & \qw
    & \cdots & & \ctrl{-3} & \qw & \ctrl{-3}
    & \ctrl{-3} & \qw & \qw & \qw
    \gategroup{5}{8}{11}{10}{1.0em}{--}
    \gategroup{2}{14}{11}{16}{1.0em}{--}
}\)
\caption{The explicit implementation of a location-controlled swap circuit. This circuit swaps the qubit $u_{\ell_t+\ell_q+1}$ from the working space \texttt{Work1} and the qubit \texttt{Sign}. Here $K^* = K - 3, k^* = k - 2$. We assume that $k \le \ell_t + \ell_q + 1 \le K$, where the parameters $k \le K$ are known in advance.}
\label{fig:ctrl_swap}
\end{figure}
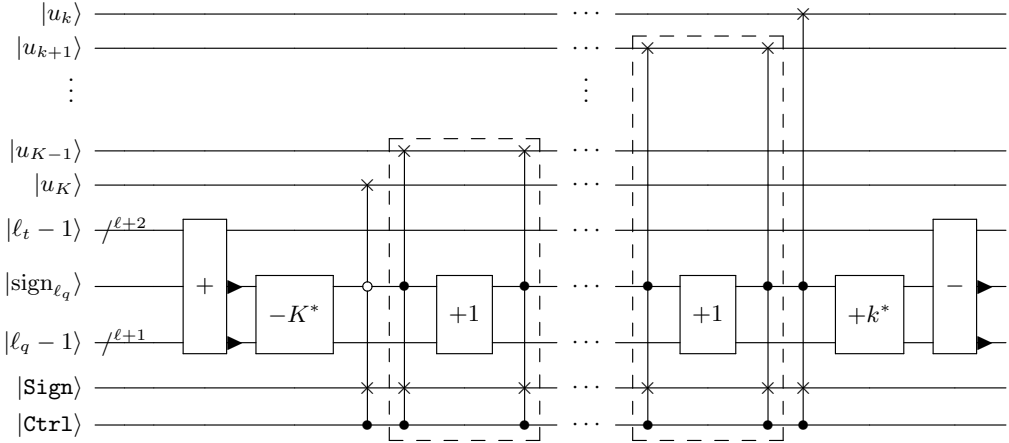

\paragraph{Location-controlled $\mathsf{Add, Sub, Swap}$ circuits.}
We first present explicit quantum circuit implementations of the location-controlled arithmetic operations appearing in Figures \ref{fig:all_circuit_1} and \ref{fig:all_circuit_2}. 
The main technical difficulty lies in realizing arithmetic operations whose execution is \emph{coherently controlled} by a location value stored in a different quantum register: 
unlike standard controlled gates, where the control qubits and active qubits are both fixed and local, here the control information specifies \emph{which positions} in the working registers should be acted upon. 
As a first step toward addressing this challenge, we begin by describing the location-controlled swap circuit, as shown in Figure \ref{fig:ctrl_swap}. 

To understand the correctness of the location-controlled swap circuit, we observe that a SWAP gate between a qubit at a specific location in \texttt{Work1} and the \texttt{Sign} qubit is executed, if and only if the two corresponding controlled-SWAP gates around the $+1$ gate (the dashed boxes in Figure \ref{fig:ctrl_swap}) do not cancel each other. 
More precisely, when $\ell_t + \ell_q + 1 = j \in [k+1, K-1]$, the two controlled-SWAP gates associated with $u_j$ differ only in the neighborhood of the $(K-j)$-th $+1$ gate acting on the $\ell_q$ register. In this case, these two controlled-SWAP gates are conditioned on different values of $\mathrm{sign}_{\ell_q}$ and therefore do not cancel each other.

We emphasize that the arithmetic gates acting on the \texttt{Length} registers (including the $+$, $-$, $-K^*$, $+k^*$, and $+1$ gates) are not controlled by the \texttt{Ctrl} wire, which represents the external control of the entire circuit block. The reason is that these arithmetic operations on the \texttt{Length} registers are eventually canceled out, regardless of whether the whole circuit block is activated by external control conditions.

\begin{figure}[ht]
\centering
\footnotesize 
\(\Qcircuit @C=0.6em @R=0.6em {
    & \lstick{\ket{0}} & \qw & \qw & \qw & \qw & \qw & \multigate{2}{\substack{\mathrm{M} \\ \mathrm{A} \\ \mathrm{J}}}& \qw & \qw & \qw & \qw & \cdots &  & \qw & \qw & \qw & \qw & \qw & \qw  & \cdots & & \qw &\qw &\multigate{2}{\substack{\mathrm{U} \\ \mathrm{M} \\ \mathrm{A}}} & \qw & \qw & \qw \\
    & \lstick{\ket{u_K}} & \qw & \qw & \qw & \qw & \qw & \ghost{\substack{\mathrm{M} \\ \mathrm{A} \\ \mathrm{J}}}& \qw & \qw & \qw & \qw & \cdots &  & \qw & \qw & \qw & \qw & \qw & \qw  & \cdots & & \qw & \qw & \ghost{\substack{\mathrm{U} \\ \mathrm{M} \\ \mathrm{A}}}  & \qw & \qw & \qw \\
    & \lstick{\ket{v_K}} & \qw & \qw & \qw & \qw & \qw & \ghost{\substack{\mathrm{M} \\ \mathrm{A} \\ \mathrm{J}}}& \ctrl{6} & \qw & \ctrl{6} & \qw & \cdots &  & \qw & \qw & \qw & \qw & \qw & \qw  & \cdots & & \qw & \qw & \ghost{\substack{\mathrm{U} \\ \mathrm{M} \\ \mathrm{A}}} & \qw & \qw & \qw\\
    & \lstick{\vdots\ } & & & & & \\
    & & & & & & \\
    & \lstick{\ket{v_{k-1}}} & \qw & \qw & \qw & \qw & \qw  & \qw & \qw & \qw  & \qw  & \qw &  \cdots & & \qw & \multigate{2}{\substack{\mathrm{M} \\ \mathrm{A} \\ \mathrm{J}}} & \qw & \multigate{2}{\substack{\mathrm{U} \\ \mathrm{M} \\ \mathrm{A}}} & \qw & \qw & \cdots & & \qw & \qw & \qw & \qw & \qw  & \qw \\
    & \lstick{\ket{u_k}} & \qw & \qw & \qw & \qw & \qw  & \qw& \qw & \qw  & \qw  & \qw &  \cdots & & \qw &  \ghost{\substack{\mathrm{M} \\ \mathrm{A} \\ \mathrm{J}}} & \qw &\ghost{\substack{\mathrm{U} \\ \mathrm{M} \\ \mathrm{A}}} & \qw & \qw & \cdots & & \qw & \qw & \qw & \qw & \qw  & \qw\\
    & \lstick{\ket{v_k}} & \qw & \qw & \qw & \qw & \qw  & \qw & \qw & \qw & \qw & \qw &  \cdots & & \qw &  \ghost{\substack{\mathrm{M} \\ \mathrm{A} \\ \mathrm{J}}} & \ctrl{1} &\ghost{\substack{\mathrm{U} \\ \mathrm{M} \\ \mathrm{A}}} & \qw & \qw & \cdots & & \qw & \qw & \qw & \qw & \qw  & \qw \\
    & \lstick{\ket{\mathrm{sign}}} & \qw & \qw & \qw & \qw & \qw  & \qw & \targ & \qw & \targ & \qw & \cdots &  & \qw & \qw & \targ & \qw & \qw & \qw & \cdots & & \qw & \qw & \qw & \qw & \qw  & \qw \\
    & \lstick{\ket{\mathrm{sign}_{\ell_s}}} & \qw & \qw & \qw & \qw  & \multigate{1}{-K_2^*} & \ctrl{-7} & \qw & \multigate{1}{-1} & \qw & \qw & \cdots & & \multigate{1}{-1} & \ctrl{-2} & \qw & \ctrl{-2} & \multigate{1}{+1} & \qw & \cdots & & \qw & \multigate{1}{+1} & \ctrl{-7} & \multigate{1}{+K_2^*} & \qw & \qw\\
    & \lstick{\ket{\ell_s-1}} & \qw & \qw\slash^{\ell + 2} & \qw & \qw & \ghost{-K_2^*} &\qw& \qw  & \ghost{-1} & \qw & \qw & \cdots &  & \ghost{-1} & \qw & \qw & \qw & \ghost{+1} & \qw & \cdots & & \qw & \ghost{+1} & \qw & \ghost{+K_2^*} & \qw & \qw\\
    & \lstick{\ket{\mathrm{sign}_{\ell_q}}} & \qw & \qw & \qw & \multigateouttwohi{2}{+} & \multigate{1}{-K_1^*} & \ctrl{-2} & \ctrl{-3} & \multigate{1}{+1} & \ctrl{-3} & \qw & \cdots &  & \multigate{1}{+1} & \ctrl{-2} & \ctrl{-3} & \ctrl{-2} & \multigate{1}{-1} & \qw & \cdots & & \qw & \multigate{1}{-1} & \ctrl{-2} & \multigate{1}{+K_1^*} & \multigateouttwohi{2}{-} & \qw \\
    & \lstick{\ket{\ell_q-1}} & \qw & \qw\slash^{\ell + 1} & \qw & \ghost{+} & \ghost{-K_1^*}  & \qw & \qw & \ghost{+1} & \qw & \qw & \cdots &  & \ghost{+1} & \qw & \qw & \qw & \ghost{-1} & \qw & \cdots & & \qw & \ghost{-1} &\qw & \ghost{+K_1^*} & \ghost{-} & \qw \\
    & \lstick{\ket{\ell_t-1}} & \qw & \qw\slash^{\ell + 2} & \qw & \ghost{+} & \qw & \qw & \qw & \qw & \qw & \qw & \cdots & & \qw & \qw & \qw & \qw & \qw & \qw  & \cdots & & \qw & \qw & \qw & \qw & \ghost{-} & \qw \\
    & \lstick{\ket{\texttt{Ctrl}}} & \qw & \qw & \qw & \qw & \qw & \ctrl{-3} & \ctrl{-3} & \qw & \ctrl{-3} & \qw & \cdots & & \qw & \ctrl{-3} & \ctrl{-3} & \ctrl{-3} & \qw & \qw  & \cdots & & \qw & \qw & \ctrl{-3} & \qw & \qw & \qw
    \gategroup{3}{9}{15}{9}{0.8em}{--}
    \gategroup{3}{11}{15}{11}{0.8em}{--}
    \gategroup{8}{17}{15}{17}{0.8em}{--}
    \gategroup{10}{2}{14}{28}{1.2em}{-}
}\)
\caption{A template for the explicit implementation of a location-controlled addition/subtraction circuit. The circuit in the figure is designed to perform the arithmetic operation $(\texttt{Sign}, r) \gets (\texttt{Sign}, r) - 2^{\ell_s} r'$.
It operates on the qubits indexed from $(\ell_t + \ell_q + 2)$ to $(n + 3 - \ell_s)$ on the working registers \texttt{Work1} and \texttt{Work2}. Here $K_1^* = K - 3$ and $K_2^* = n + 3 - K$. We assume that $k \le \ell_t + \ell_q + 2$ and $n + 3 - \ell_s \le K$, where the parameters $k \le K$ are known in advance. 
For the location-controlled addition circuit implementing $r \gets r + 2^{\ell_s} r'$ without activating the \texttt{Sign} register, the Toffoli gates in the dashed boxes can be removed.
For the location-controlled addition/subtraction circuits acting on $t$'s, the control is one-sided. Therefore, half of the $\pm K^*, \pm 1$ gates inside the solid box, which are used to implement the location-control mechanism, can be omitted.
}
\label{fig:ctrl_add}
\end{figure}
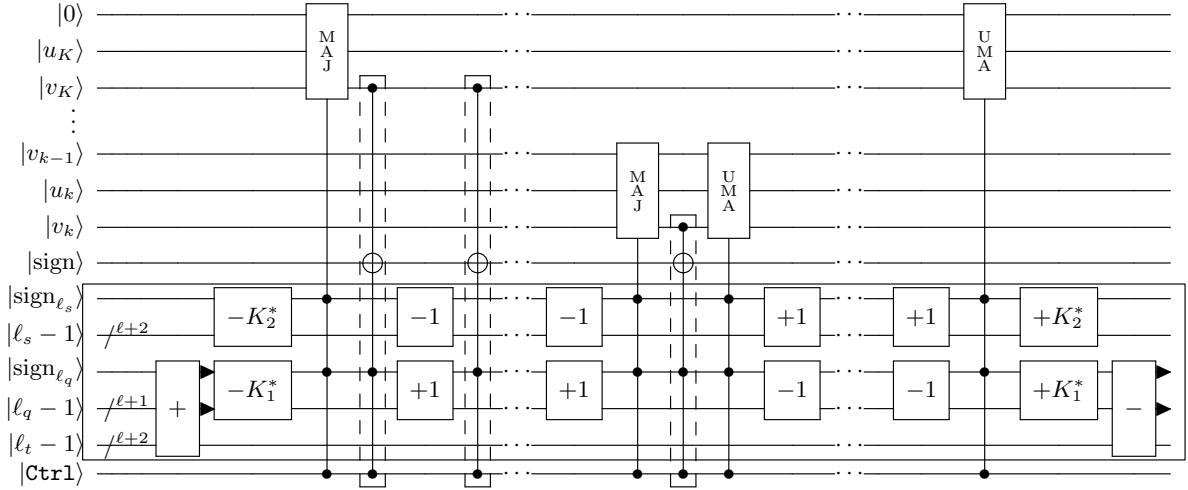

Keeping the implementation principle of the location-controlled swap gate in mind, we next present explicit circuit diagrams for the location-controlled addition/subtraction circuit blocks, adopting the ripple-carry adder using MAJ and UMA blocks \cite{cuccaro2004new}, as shown in Figure \ref{fig:ctrl_add}. 

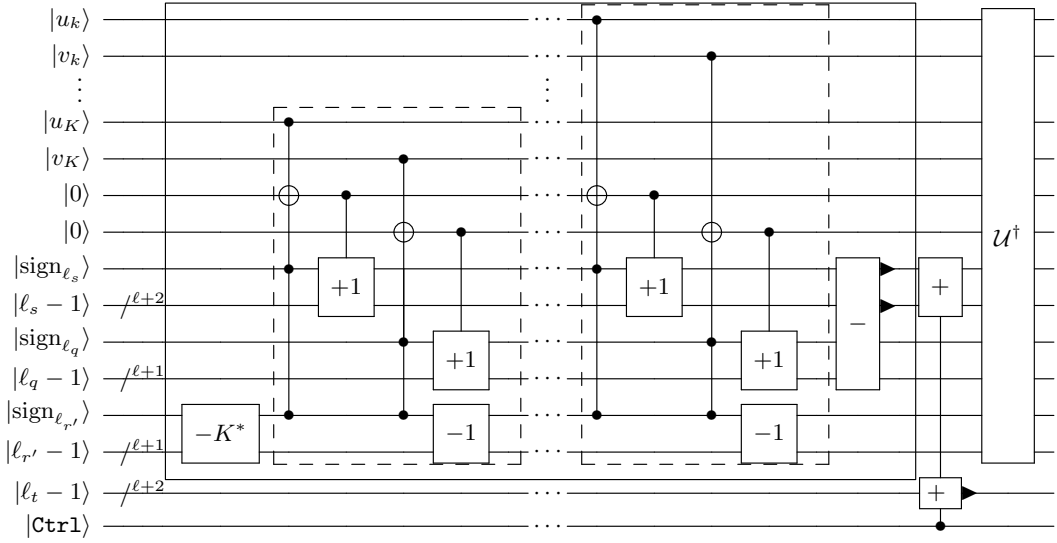
\begin{figure}[ht]
\centering
\footnotesize 
\(\Qcircuit @C=0.8em @R=0.6em {
    & \lstick{\ket{u_k}} & \qw & \qw & \qw & \qw 
    & \qw & \qw & \qw & \qw & \qw 
    & \qw & \cdots & & \ctrl{6} 
    & \qw & \qw & \qw & \qw & \qw & \qw
    & \qw & \multigate{13}{\mathcal{U}^{\dagger}} & \qw \\
    & \lstick{\ket{v_k}} & \qw & \qw & \qw & \qw 
    & \qw & \qw & \qw & \qw & \qw 
    & \qw & \cdots & & \qw 
    & \qw & \ctrl{6} & \qw & \qw & \qw & \qw
    & \qw & \ghost{\mathcal{U}^{\dagger}} & \qw \\
    & \lstick{\vdots\ } & & & & & & & & & 
    & & \vdots & & & & & & & & & & & \\
    & & & & & & & & & & 
    & & & & & & & & & & & & & \\
    & \lstick{\ket{u_K}} & \qw & \qw & \qw & \qw 
    & \ctrl{2} & \qw & \qw & \qw & \qw 
    & \qw & \cdots & & \qw 
    & \qw & \qw & \qw & \qw & \qw & \qw
    & \qw & \ghost{\mathcal{U}^{\dagger}} & \qw \\
    & \lstick{\ket{v_K}} & \qw & \qw & \qw & \qw 
    & \qw & \qw & \ctrl{2} & \qw & \qw 
    & \qw & \cdots & & \qw 
    & \qw & \qw & \qw & \qw & \qw & \qw
    & \qw & \ghost{\mathcal{U}^{\dagger}} & \qw \\
    & \lstick{\ket{0}} & \qw & \qw & \qw & \qw 
    & \targ & \ctrl{2} & \qw & \qw & \qw 
    & \qw & \cdots & & \targ 
    & \ctrl{2} & \qw & \qw & \qw & \qw & \qw
    & \qw & \ghost{\mathcal{U}^{\dagger}} & \qw \\
    & \lstick{\ket{0}} & \qw & \qw & \qw & \qw 
    & \qw & \qw & \targ & \ctrl{3} & \qw 
    & \qw & \cdots & & \qw 
    & \qw & \targ & \ctrl{3} & \qw & \qw & \qw
    & \qw & \ghost{\mathcal{U}^{\dagger}} & \qw \\
    & \lstick{\ket{\mathrm{sign}_{\ell_s}}} & \qw & \qw & \qw & \qw 
    & \ctrl{-2} & \multigate{1}{+1} & \qw & \qw & \qw 
    & \qw & \cdots & & \ctrl{-2} 
    & \multigate{1}{+1} & \qw & \qw & \qw & \multigateouttwohi{3}{-} & \qw 
    & \multigate{1}{+} & \ghost{\mathcal{U}^{\dagger}} & \qw \\
    & \lstick{\ket{\ell_s-1}} & \qw & \qw\slash^{\ell + 2} & \qw & \qw 
    & \qw & \ghost{+1} & \qw & \qw & \qw 
    & \qw & \cdots & & \qw 
    & \ghost{+1} & \qw & \qw & \qw & \ghost{-} & \qw 
    & \ghost{+}\qwx[5] & \ghost{\mathcal{U}^{\dagger}} & \qw \\
    & \lstick{\ket{\mathrm{sign}_{\ell_q}}} & \qw & \qw & \qw & \qw 
    & \qw & \qw & \ctrl{-3} & \multigate{1}{+1} & \qw 
    & \qw & \cdots & & \qw 
    & \qw & \ctrl{-3} & \multigate{1}{+1} & \qw & \ghost{-} & \qw 
    & \qw & \ghost{\mathcal{U}^{\dagger}} & \qw \\
    & \lstick{\ket{\ell_q-1}} & \qw & \qw\slash^{\ell + 1} & \qw & \qw 
    & \qw & \qw & \qw & \ghost{+1} & \qw 
    & \qw & \cdots & & \qw 
    & \qw & \qw & \ghost{+1} & \qw & \ghost{-} & \qw 
    & \qw & \ghost{\mathcal{U}^{\dagger}} & \qw \\
    & \lstick{\ket{\mathrm{sign}_{\ell_{r'}}}} & \qw & \qw & \qw & \multigate{1}{-K^*} 
    & \ctrl{-4} & \qw & \ctrl{-4} & \multigate{1}{-1} & \qw 
    & \qw & \cdots & & \ctrl{-4} 
    & \qw & \ctrl{-2} & \multigate{1}{-1} & \qw & \qw & \qw
    & \qw & \ghost{\mathcal{U}^{\dagger}} & \qw \\
    & \lstick{\ket{\ell_{r'} - 1}} & \qw & \qw\slash^{\ell + 1} & \qw & \ghost{-K^*} 
    & \qw & \qw & \qw & \ghost{-1} & \qw 
    & \qw & \cdots & & \qw 
    & \qw & \qw & \ghost{-1} & \qw & \qw & \qw
    & \qw & \ghost{\mathcal{U}^{\dagger}} & \qw \\
    & \lstick{\ket{\ell_t - 1}} & \qw & \qw\slash^{\ell + 2} & \qw & \qw
    & \qw & \qw & \qw & \qw & \qw 
    & \qw & \cdots & & \qw 
    & \qw & \qw & \qw & \qw & \qw & \qw
    & \gateout{+} & \qw & \qw \\
    & \lstick{\ket{\texttt{Ctrl}}} & \qw & \qw & \qw & \qw 
    & \qw & \qw & \qw & \qw & \qw 
    & \qw & \cdots & & \qw 
    & \qw & \qw & \qw & \qw & \qw & \qw
    & \ctrl{-1} & \qw & \qw
    \gategroup{1}{6}{14}{21}{1.35em}{-}
    \gategroup{5}{7}{14}{11}{1.0em}{--}
    \gategroup{1}{15}{14}{19}{1.0em}{--}
}\)
\caption{The explicit implementation of the length updating circuit. 
This circuit updates $\ell_t$ to the bit length of the new $t$.
Here, $K^* = n + 3 - K$, and $\mathcal{U}$ denotes the collection of quantum operations in the outer solid box.
The bit-wise operations shown in the dashed boxes are applied independently to each qubit pair $\ket{u_i, v_i}, i = K, K - 1, \ldots, k$.
We assume that the previous value of $\ell_t$ does not exceed $k$, and that $\ell_{r'} \ge n + 3 - K$, i.e., the updated value of $\ell_t$ is guaranteed to be at most $K$, where the parameters $k \le K$ are fixed and known in advance.}
\label{fig:len_update}
\end{figure}

\paragraph{Length-update circuits.}
Finally, we describe the length-update circuits used at the end of each EEA iteration to update the values of $\ell_t$ and $\ell_{r'}$. We focus on the circuit that updates $\ell_t$ to the length of the new value of $t$, as shown in Figure \ref{fig:len_update}; the circuit for updating $\ell_{r'}$ is similar and therefore omitted.

To see why this circuit correctly updates the length of $t$, it is helpful to first consider a simplified version in which the control on $\ell_{r'}$ is removed. Inside the dashed box in Figure \ref{fig:len_update}, once the circuit encounters a position $j$ such that $u_j = 1$, the first auxiliary qubit is irreversibly set to $1$, which in turn causes the $+1$ gate to be applied from that point onward. Similarly, once a position with $v_j = 1$ is encountered, the $+1$ gate is again permanently activated.

As a result, after the loop within the dashed box is completed, the values stored in the registers $\ell_s$ and $\ell_q$ differ by exactly the amount equal to the difference between the new and the original lengths of $t$. This difference is then added to the register $\ell_t$, yielding the correct updated length. Finally, the operation $\mathcal{U}^\dagger$ uncomputes all auxiliary computations in the outer solid box, restoring the auxiliary registers to their original states.

%% file: sec_circuit/sub_active.tex
\subsection{Step-dependent active windows}
\label{subsec:actwindow}

To reduce the quantum gate complexity for implementing our space-efficient EEA, we introduce step-dependent active windows for each location-controlled operation. As illustrated in the figures of the previous subsection, we define $k$ and $K$ as known bounds on the active indices of the \texttt{Work} registers.

For instance, at step $T$, a location-controlled Swap circuit acts on the $(\ell_t + \ell_q + 1)$-th qubit of the \texttt{Work1} register. Over all possible inputs $x$ used to compute the modular inverse $x^{-1} \bmod q$, the quantity $\ell_t + \ell_q + 1$ is bounded below by $k := k(T)$ and above by $K := K(T)$. Therefore, applying location-controlled operations to qubits in the \texttt{Work1} register with indices outside the interval $[k, K]$ is unnecessary. We refer to this interval as the \emph{active window}.

We next specify the step-dependent active windows for each type of location-controlled operation. All bounds are taken over all \emph{possible inputs} $x$, namely, those inputs $x$ for which the location-controlled circuits are activated by the outer control qubits. The constant $c$ equals to $1 / \log_2\left(\frac{\sqrt{5} + 1}{2}\right)$. We defer the proofs to Appendix \ref{app:actwindow}.

\begin{itemize}

\item \textbf{Location-controlled addition/subtraction on $r$'s.}  
These operations act on qubits indexed from $(\ell_t + \ell_q + 2)$ to $(n + 3 - \ell_s)$. 
The upper bound $K_1 = K_1(T) = n + 3$; the lower bound $k_1 = k_1(T)$ is defined by
\[
    \ell_t + \ell_q + 2 \ge k_1(T)
    := \max\left\{ \ceil{\frac{T - n - 2}{4c - 1}}, 1 \right\} + 2.
\]

\item \textbf{Location-controlled swap.} 
These operations act on the $(\ell_t + \ell_q + 1)$-th qubit. 
The two-sided bounds are defined by
\[
\begin{split}
    \ell_t + \ell_q + 1 
    & \ge k_2(T) := \max\left\{\ceil{\frac{T - 3(n + 2)}{4c - 3}}, 1\right\} + 1, \\
    \ell_t + \ell_q + 1 
    & \le K_2(T) := \min\left\{\floor{T/2} + 2, n + 2\right\}.
\end{split}
\]

\item \textbf{Location-controlled addition/subtraction on $t$'s.}  
These operations act on the first $(\ell_t + 1)$ qubits. 
This is a one-sided controlled operation, the upper bound $K_3(T)$ is defined by
\[
    \ell_t + 1 \le K_3(T) 
    := \min\left\{\ceil{T/4} + 1, n + 1\right\}.
\]

\item \textbf{Length-update circuit for updating $\ell_t$.}  
We execute the length-update circuit only at step indices $T$ that are multiples of four.
This circuit acts on qubits indexed from $\ell_t$ to $(n + 3 - \ell_{r'})$, and the corresponding two-sided active window is defined by
\[
\begin{split}
    \ell_t 
    & \ge k_4(T) := \max\left\{\ceil{\frac{T - 4(n + 2)}{4c - 4}}, 1\right\}, \\
    n + 3 - \ell_{r'} 
    & \le K_4(T) := \min\left\{T/4 + 3, n + 3\right\}.
\end{split}
\]

\item \textbf{Length-update circuit for updating $\ell_{r'}$.}  
Let $\ell_{t}^*, \ell_{r'}^*$ denote the updated value of $\ell_t, \ell_{r'}$ after the length update. 
This circuit acts on qubits indexed from $\ell_t^* + 2$ to $(n + 4 - \ell_{r'}^*)$, and the corresponding two-sided active window is defined by
\[
\begin{split}
    \ell_t^* + 2 
    & \ge k_5(T) := \ceil{T/(4c)}, \\
    n + 4 - \ell_{r'}^* 
    & \le K_5(T) := \min\left\{T/4 + 4, n + 3\right\}.
\end{split}
\]

\end{itemize}

%% file: sec_resource/sec_resource.tex
\section{Explicit Quantum Circuit Construction and Resource Estimation}
\label{sec:resource_estimation}

We adopt the resource-estimation methodology introduced by Roetteler et al.~\cite{roetteler2017quantum}.
In Shor’s algorithm for the ECDLP, the dominant quantum cost arises from the sequence of controlled elliptic-curve point additions performed in superposition. 
We consider the point additions in affine coordinates over the finite field $\mathbb{F}_p$, where $p$ is an $n$-bit prime.
At this level, point addition reduces to a fixed sequence of modular additions, subtractions, doublings, multiplications, squarings, and inversions.
Among these operations, modular inversion typically dominates both the required circuit width and the Toffoli gate count.

\input{sec_resource/sub_space}
\input{sec_resource/sub_asymp}
\input{sec_resource/sub_num}

\input{sec_resource/sub_ecdlp}

%% file: sec_resource/sub_space.tex
\subsection{Space complexity improvement from modular inversion}

In \cite{roetteler2017quantum}, their implementation of controlled point addition requires
\[
    9n + 2\lceil \log_2 n\rceil + 10
\]
logical qubits in total. Within this construction, the inversion subroutine itself uses
\[
    7n + 2\lceil \log_2 n\rceil + 9
\]
logical qubits.
The remaining qubits consist of a single global control qubit and two additional $n$-qubit registers that are used to store intermediate values during each inversion.

Subsequent work by H\"{a}ner et al.~\cite{haner2020improved} further optimized the trade-offs between Toffoli count, circuit depth, and width in point-addition circuits.
Under a low-width design, they report a reduction in the number of logical qubits for a 256-bit curve from $2338$ to $2124$, together with substantial improvements in both $T$-count and $T$-depth.

Our contribution is orthogonal to these optimizations.
Rather than exploring trade-offs between Toffoli count, circuit depth, and qubit width, we focus exclusively on reducing the qubit width required to implement modular inversion.
We present a space-efficient and exact reversible modular inversion procedure based on the register-sharing technique originally proposed by Proos and Zalka \cite{proos2003shor}.
This approach reduces the qubit width of modular inversion to
\[
    3n + 4\floor{\log_2 n} + 20,
\]
which is also summarized in Table \ref{tab:mod_arith_costs}.

%% file: sec_resource/sub_asymp.tex
\subsection{Asymptotic gate count for modular inversion}
\label{subsec:inversion_costs}

The leading term of the Toffoli/CNOT gate count arises from the implementation of location-controlled operations, which require $O(n^2 \log_2 n)$ Toffoli/CNOT gates. Within the detailed circuit construction of these operations, the leading-order cost arises from arithmetic operations on the \texttt{Length} registers; the remaining components contribute only $O(n^2)$ Toffoli/CNOT gates.

As shown in Appendix \ref{app:adder}, an $\ell$-bit quantum incrementer requires $(2\ell - 2)$ Toffoli gates and $(\ell + 2)$ CNOT gates. Using the step-dependent active window described in Section \ref{subsec:actwindow}, we derive the asymptotic Toffoli gate count associated with each type of location-controlled operation. Recall that $c = 1 / \log_2\left(\frac{\sqrt{5} + 1}{2}\right)$.
\begin{itemize}
\item \textbf{Location-controlled addition/subtraction on $r$'s.}  
At step $T$, these operations consist of $4\bigl(K_1(T) - k_1(T)\bigr) + O(1)$ quantum incrementers. The total Toffoli gate count is
\[
    2\sum_{T=1}^{4\ceil{cn}} 8\log_2 n \bigl(K_1(T) - k_1(T)\bigr) + O(n^2)
    = (32c + 8)n^2\log_2 n + O(n^2).
\]

\item \textbf{Location-controlled swap.}  
At step $T$, these operations consist of $\bigl(K_2(T) - k_2(T)\bigr) + O(1)$ quantum incrementers. The resulting Toffoli gate count is
\[
    \sum_{T=1}^{4\ceil{cn}} 2\log_2 n \bigl(K_2(T) - k_2(T)\bigr) + O(n^2)
    = (4c + 1)n^2\log_2 n + O(n^2).
\]

\item \textbf{Location-controlled addition/subtraction on $t$'s.}  
At step $T$, these operations consist of $2\bigl(K_3(T) - k_3(T)\bigr) + O(1)$ quantum incrementers. The corresponding Toffoli gate count is
\[
    2\sum_{T=1}^{4\ceil{cn}} 4\log_2 n \bigl(K_3(T) - k_3(T)\bigr) + O(n^2)
    = (32c - 16)n^2\log_2 n + O(n^2).
\]

\item \textbf{Length-update circuit for updating $\ell_t$.}  
We execute the length-update circuit only at step indices $T$ that are multiples of four. At step $T = 4T'$, these operations consist of
$6\bigl(K_4(4T') - k_4(4T')\bigr) + O(1)$ quantum incrementers. The total Toffoli gate count is
\[
    \sum_{T'=1}^{\ceil{cn}} 12\log_2 n \bigl(K_4(4T') - k_4(4T')\bigr) + O(n^2)
    = 6c n^2\log_2 n + O(n^2).
\]

\item \textbf{Length-update circuit for updating $\ell_{r'}$.}  
We execute the length-update circuit only at step indices $T$ that are multiples of four. At step $T = 4T'$, these operations consist of
$6\bigl(K_5(4T') - k_5(4T')\bigr) + O(1)$ quantum incrementers. The resulting Toffoli gate count is
\[
    \sum_{T'=1}^{\ceil{cn}} 12\log_2 n \bigl(K_5(4T') - k_5(4T')\bigr) + O(n^2)
    = (6c - 6)n^2\log_2 n + O(n^2).
\]
\end{itemize}
Combining all contributions, the total number of Toffoli gates required by our modular inversion circuit grows asymptotically as
\[
    2(80c - 13)n^2\log_2 n + O(n^2)
    \approx 204 n^2 \log_2 n + O(n^2).
\]
The asymptotical bound of CNOT gates required by our modular inversion circuit is therefore about $102n^2\log_2 n + O(n^2)$.

%% file: sec_resource/sub_num.tex
\subsection{Numerical experiments for modular inversion}
All resource-estimation experiments were conducted on a Linux server equipped with an AMD EPYC 7302 processor (up to 3.0\,GHz). Circuit generation and compilation were performed using IBM Qiskit. The constructed circuits were transpiled into a gate basis consisting of \texttt{CCX}, \texttt{CX}, and \texttt{X} gates, in order to obtain accurate Toffoli-gate and CNOT-gate counts.

Our implementation is built directly from the explicit circuit construction presented in Section~\ref{sec:circuit}. Following the analysis of the one-step modular-inversion circuit, we implemented each circuit component in Qiskit as an independent module, including the pre-shift and post-shift blocks, the location-controlled arithmetic and swap blocks, the phase-update logic, and the length-update circuits. These modules were then synthesized into full modular-inversion circuits by composing them according to the step schedule of 
Algorithm~\ref{alg:opt} and the corresponding step-dependent active windows.

The corresponding circuit implementation is included in our open-source codebase.\footnote{\label{footnote:repo}GitHub repository: \href{https://github.com/ZeroWang030221/Space-Efficient-Quantum-Algorithm-for-Elliptic-Curve-Discrete-Logarithms-with-Resource-Estimation}{https://github.com/ZeroWang030221/Space-Efficient-Quantum-Algorithm-for-Elliptic-Curve-Discrete-Logarithms-with-Resource-Estimation}} For each problem size $n$, we generated the full circuit instance up to the theoretical upper bound on the number of cycles derived in this work, transpiled it into the \texttt{CCX}, \texttt{CX}, and \texttt{X} basis, and recorded the resulting gate counts. We report results for $n \in \{64,128,160,192,224,256,384,512\}$. The numerical values are listed in Table~\ref{tab:modinv_counts}. 

It is noteworthy that our Toffoli gate count exceeds the CNOT count in Table~\ref{tab:modinv_counts}. This ratio arises because most arithmetic sub-components are nested within two external control bits (Phase 1 and Phase 2 in Section~\ref{subsec:4phase}) that transform internal CNOT gates to Toffoli gates. Additionally, our constant adders are optimized for increment operations, significantly reducing the CNOT overhead compared to standard constructions (see Appendix \ref{app:adder} for details).

\begin{table}[htbp]
\centering
% \label{tab:modinv_counts}
\begin{tabular}{c|cc}
\hline
$n$ & Toffoli ($\times 10^8$) & CNOT ($\times 10^8$) \\
\hline
64  & 0.10 & 0.07 \\
128 & 0.44 & 0.32 \\
160 & 0.78 & 0.54 \\
192 & 1.12 & 0.77 \\
224 & 1.51 & 1.04 \\
256 & 1.97 & 1.36 \\
384 & 3.53 & 3.28 \\
512 & 6.24 & 5.82 \\
\hline
\end{tabular}
\caption{Toffoli and CNOT gate counts for modular inversion circuits over $n$-bits prime fields.}
\label{tab:modinv_counts}
\end{table}

%% file: sec_resource/sub_ecdlp.tex
\subsection{Resource Estimates for ECDLP}
\label{subsec:component_costs}

We next propagate the improvement of Section \ref{subsec:inversion_costs} from the modular-inversion subroutine to the full quantum circuit for solving the ECDLP.
To keep the accounting uniform, we use only \emph{Toffoli}-based cost estimates in this subsection. More precisely, we adopt the high-level affine-Weierstrass point-addition decomposition of \cite[Figure 10]{roetteler2017quantum} and \cite[Figure 9]{haner2020improved}, and evaluate all arithmetic components in Toffoli gates using the resource table of \cite[Table 1]{roetteler2017quantum}. 

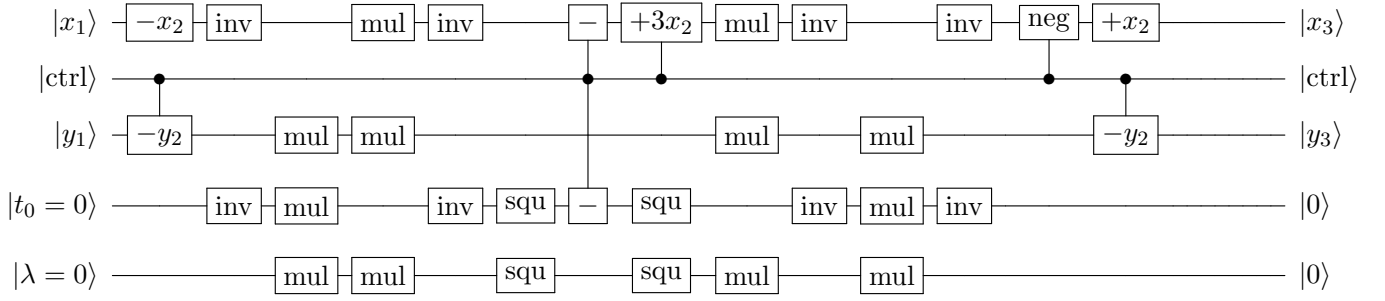
\begin{figure}[t]
\centering
\hspace*{1.0cm}%
\resizebox{0.9\linewidth}{!}{%
$\Qcircuit @C=0.5em @R=1.2em {
\lstick{\ket{x_1}}
    & \gate{-x_2}
    & \gate{\mathrm{inv}}
    & \qw
    & \gate{\mathrm{mul}}
    & \gate{\mathrm{inv}}
    & \qw
    & \gate{-}
    & \gate{+3x_2}
    & \gate{\mathrm{mul}}
    & \gate{\mathrm{inv}}
    & \qw
    & \gate{\mathrm{inv}}
    & \qw
    & \gate{\mathrm{neg}}
    & \gate{+x_2}
    & \qw
    & \qw
    & \qw
    & \qw
    & \qw
    & \qw
    & \qw
    & \qw
    & \rstick{\ket{x_3}} \qw \\
\lstick{\ket{\mathrm{ctrl}}}
    & \ctrl{1}
    & \qw
    & \qw
    & \qw
    & \qw
    & \qw
    & \ctrl{-1} \qwx[2]
    & \ctrl{-1}
    & \qw
    & \qw
    & \qw
    & \qw
    & \qw
    & \ctrl{-1}
    & \ctrl{1}
    & \qw
    & \qw
    & \qw
    & \qw
    & \qw
    & \qw
    & \qw
    & \qw
    & \rstick{\ket{\mathrm{ctrl}}} \qw \\
\lstick{\ket{y_1}}
    & \gate{-y_2}
    & \qw
    & \gate{\mathrm{mul}}
    & \gate{\mathrm{mul}}
    & \qw
    & \qw
    & \qw
    & \qw
    & \gate{\mathrm{mul}}
    & \qw
    & \gate{\mathrm{mul}}
    & \qw
    & \qw
    & \qw
    & \gate{-y_2}
    & \qw
    & \qw
    & \qw
    & \qw
    & \qw
    & \qw
    & \qw
    & \qw
    & \rstick{\ket{y_3}} \qw \\
\lstick{\ket{t_0=0}}
    & \qw
    & \gate{\mathrm{inv}}
    & \gate{\mathrm{mul}}
    & \qw
    & \gate{\mathrm{inv}}
    & \gate{\mathrm{squ}}
    & \gate{-}
    & \gate{\mathrm{squ}}
    & \qw
    & \gate{\mathrm{inv}}
    & \gate{\mathrm{mul}}
    & \gate{\mathrm{inv}}
    & \qw
    & \qw
    & \qw
    & \qw
    & \qw
    & \qw
    & \qw
    & \qw
    & \qw
    & \qw
    & \qw
    & \rstick{\ket{0}} \qw \\
\lstick{\ket{\lambda=0}}
    & \qw
    & \qw
    & \gate{\mathrm{mul}}
    & \gate{\mathrm{mul}}
    & \qw
    & \gate{\mathrm{squ}}
    & \qw
    & \gate{\mathrm{squ}}
    & \gate{\mathrm{mul}}
    & \qw
    & \gate{\mathrm{mul}}
    & \qw
    & \qw
    & \qw
    & \qw
    & \qw
    & \qw
    & \qw
    & \qw
    & \qw
    & \qw
    & \qw
    & \qw
    & \rstick{\ket{0}} \qw
    &
}$}
\caption{
Schematic low-width controlled affine Weierstrass point-addition circuit, adapted from the construction of  \cite[Figure 10]{roetteler2017quantum}.
}
\label{fig:loww_point_addition}
\end{figure}

\paragraph{Controlled affine point addition.}
As discussed in the preliminaries, given affine points $P_1=(x_1,y_1)$ and $P_2=(x_2,y_2)$ that are not inverse to each other, one computes
\[
    x_3=\lambda^2-x_1-x_2,\qquad
    y_3=\lambda(x_1-x_3)-y_1,
\]
so that $(x_3,y_3)=P_1+P_2$.
The corresponding controlled circuit is shown schematically in Figure \ref{fig:loww_point_addition}. Following \cite[Figure 10]{roetteler2017quantum} and \cite[Figure 9]{haner2020improved}, we count one such point addition as consisting of 4 modular divisions, 4 modular multiplications, 1 modular squaring, and a constant number of modular additions/subtractions/negations.
The latter contribute only $O(n\log_2 n)$ Toffoli gates in total and are therefore absorbed into the lower-order term.

Table \ref{tab:mod_arith_costs} shows the modular-arithmetic summary table of \cite[Table 1]{roetteler2017quantum}, but we
(i) report only the \texttt{dbl/add} variants for multiplication/squaring to reduce space overhead, and (ii) replace \texttt{inv\_modp} by our inversion module. The non-inversion Toffoli counts are taken from \cite{roetteler2017quantum}.

\begin{table}[ht]
\centering
\renewcommand{\arraystretch}{1.2}
\begin{tabular}{|l|c|c|c|}
\hline
\multirow{2}{*}{Modular arithmetic circuit} &
\multicolumn{2}{c|}{\# of logical qubits} &
\multirow{2}{*}{\# Toffoli gates} \\
\cline{2-3}
& total & ancillas & \\
\hline
\shortstack[l]{\texttt{add\_const\_modp},\\ \texttt{sub\_const\_modp}}
& $2n$ & $n$ & $16n\log_2(n) - 26.9n$ \\
\hline
\shortstack[l]{\texttt{ctrl\_add\_const\_modp},\\ \texttt{ctrl\_sub\_const\_modp}}
& $2n+1$ & $n$ & $16n\log_2(n) - 26.9n$ \\
\hline
\texttt{ctrl\_sub\_modp}
& $2n+4$ & $3$ & $16n\log_2(n) - 23.8n$ \\
\hline
\texttt{ctrl\_neg\_modp}
& $n+3$ & $2$ & $8n\log_2(n) - 14.5n$ \\
\hline
\shortstack[l]{\texttt{mul\_modp}\\(\texttt{dbl/add})}
& $3n+2$ & $2$ & $32n^2\log_2(n) - 59.4n^2$ \\
\hline
\shortstack[l]{\texttt{squ\_modp}\\(\texttt{dbl/add})}
& $2n+3$ & $3$ & $32n^2\log_2(n) - 59.4n^2$ \\
\hline
\shortstack[l]{\texttt{inv\_modp}\\(\textbf{this work})}
& $3n + 4\lfloor\log_2(n)\rfloor + O(1)$
& $n + 4\lfloor\log_2(n)\rfloor + O(1)$
& $204n^2\log_2 n$ \\
\hline
\end{tabular}
\caption{Modular-arithmetic resource summary for affine point addition over $\mathbb{F}_p$ with $n$-bit prime $p$.
The first two rows use one/two clean ancillas and may additionally borrow dirty ancillas as in \cite{roetteler2017quantum}.
All other ancillas are assumed clean and are returned to $\ket{0}$. Non-inversion entries follow \cite{roetteler2017quantum};
the inversion entry is replaced by our circuit.}
\label{tab:mod_arith_costs}
\end{table}

Therefore, a single controlled affine point addition requires
\[
    (4\cdot 204 + 4\cdot 32 + 32)n^2\log_2 n + O(n^2) = 976n^2\log_2 n + O(n^2)
\]
Toffoli gates.

\paragraph{Resource estimates for Shor's ECDLP circuit.}
For the full ECDLP attack, we do not repeat the controlled point-addition circuit
bit-by-bit for all $2n$ control bits. Instead, we follow the signed-window technique
of~\cite[Section~5.1]{haner2020improved}. Namely, we partition the
$2n$ control bits arising in the double-scalar multiplication into windows of size $w$,
and implement each window by one signed windowed point addition.

In the signed-window construction, one of the $w$ bits is used as a sign bit, while the
remaining $w-1$ bits address a precomputed table of point multiples. As shown schematically
in Fig.~10 of~\cite{haner2020improved}, one such windowed point addition consists of the same
dominant arithmetic core as a single affine Weierstrass point addition, together with six
table look-ups and one modular negation. Since the dominant $n^2\log_2 n$ term in our
Toffoli accounting comes from modular divisions, multiplications and squaring, replacing the
classically known addend by a looked-up cache point affects only lower-order terms.

To obtain a clean asymptotic bound, we choose window size $\omega = 2\log_2 n$ to fit the signed windowed point addition. A sequential quantum look-up over a table of size $2^{w-1}$ contributes $O(2^w) = O(n^2)$ Toffoli gates, and hence the six look-ups together contribute only $O(n^2)$ Toffoli gates. 
The full double-scalar multiplication now uses approximately $2n/w$ such windows, so the dominant Toffoli count of Shor's ECDLP circuit becomes
\[
    \frac{2n}{\omega}(976n^2\log n + O(n^2)) = 976n^3 + O\left(\frac{n^3}{\log_2 n}\right).
\]

Finally, we note that compression techniques and multi-run tradeoffs \cite{ekeraa2019revisiting} could potentially yield further improvements to the constant factors in our overall resource estimates.

%% file: app_proof/app_proof.tex
\section{Proof Details}

\input{app_proof/sub_stepnum}
\input{app_proof/sub_active}

%% file: app_proof/sub_stepnum.tex
\subsection{Bounds on the total number of steps}
\label{app:step_num}

Suppose that after $k$ iterations of the Extended Euclidean Algorithm, we obtain $r_{k-1} = 1$ and $r_k = 0$, with intermediate quotients $q_1, q_2, \ldots, q_{k-1}$, where $q_i = \lfloor r_{i-1} / r_i \rfloor$ for $i = 1, 2, \ldots, k-1$. Then the total number of steps required in our four-phase algorithm can be expressed as
\[
    N = 4\sum_{i=1}^{k-1} \left(\lfloor \log_2 q_i \rfloor + 1\right).
\]

The lower bound of $N$ can be derived directly. Since $r_{i-1} = q_i r_i + r_{i+1} < (q_i + 1)r_i$, it follows that
\[
    N \ge 4\sum_{i=1}^{k-1}\log_2(q_i + 1) > 4\sum_{i=1}^{k-1}\log_2\frac{r_{i-1}}{r_i} = 4\log_2 p \ge 4(n - 1).
\]
Because $N$ must be a multiple of $4$, we conclude that $N \ge 4n$.

To derive the upper bound of $N$, we take a complementary viewpoint. Instead of fixing $p$ and computing $N$, we fix $N$ and seek the smallest possible value of $p$ as $k$ and the quotient sequence $\{q_i\}_{i=1}^{k-1}$ vary. Note that $p = r_0$ can be reconstructed recursively from the sequence defined by $r_k = 0$, $r_{k-1} = 1$, and $r_{i-1} = r_{i+1} + q_i r_i, \quad i = k-1, k-2, \cdots, 1$. To minimize $p$ under this recurrence, we apply the following sequence of adjustment steps, keeping $N = 4\sum_{i=1}^{k-1}(b_i + 1)$ fixed:
\begin{itemize}
    \item We first adjust all quotients to powers of two, i.e., $q_i = 2^{b_i}$ with $b_i = \lfloor \log_2 q_i \rfloor$. This modification is justified because reducing any quotient strictly decreases the corresponding sequence $r_{i-1}, \ldots, r_0$.

    \item For any $i = 1, 2, \ldots, k-2$, if $b_i \ge 2$ for $i = 1$ and $b_i \ge 1$ for $i > 1$, we can split the EEA iteration with quotient $q_i = 2^{b_i}$ into two successive iterations with quotients $q_{i,1} = 1$ and $q_{i,2} = 2^{b_i - 1}$. The updated value of $r_{i-1}$ is then
    \[
        r_{i-1}' = r_i + q_{i,1}(r_{i+1} + q_{i,2}r_i) = r_{i+1} + (2^{b_i-1} + 1)r_i \le r_{i+1} + 2^{b_i}r_i = r_{i-1},
    \]
    implying that the modified sequence yields a smaller $p$. The condition $b_1 \ge 2$ when $i = 1$ is required since the algorithm assumes $x < p/2$.

    \item If $b_{k-1} \ge 2$, we further refine the final iteration with $q_{k-1} = 2^{b_{k-1}}$ by splitting it into two iterations with quotients $q_{k-1,1} = 2$ and $q_{k-1,2} = 2^{b_{k-1}-2}$. This adjustment differs from the previous one because $r_{k-2}$ must remain greater than $1$. The updated value of $r_{k-2}$ becomes
    \[
        r_{k-2}' = 1 + q_{k-1,1} \cdot q_{k-1,2} = 1 + 2^{b_{k-1}-1} < 2^{b_{k-1}} = r_{k-2},
    \]
    showing once again that the adjustment leads to a smaller $p$ for the same total number of steps.
\end{itemize}

After applying all possible adjustments, we obtain a quotient sequence of form $\{q_i\} = \{2, 1, 1, \ldots, 1, 2\}$ with length $k - 1 = N/4 - 2$. The corresponding sequence of $r_i$ satisfies
\[
    r_{k-1} = 1, \quad r_{k-2} = 2, \quad r_{i-1} = r_{i+1} + r_i \text{ for } i = k-2, \ldots, 2, \quad \text{and } r_0 = r_2 + 2r_1.
\]
This recurrence defines a Fibonacci sequence with $r_{k-i} = F_{i+1}$ for $i = 1, 2, \cdots, k-1$, and the minimal possible value of $p$ is thus $p = r_0 = F_{k-1} + 2F_k = F_{k+2} = F_{N/4+1}$. Together with the condition $2^{n-1} \le p < 2^n$, we obtain
\[
    \frac{1}{\sqrt{5}}\left(\frac{\sqrt{5} + 1}{2}\right)^{N/4+1} \le F_{N/4+1} + 1 \le p + 1 \le 2^n.
\]
Hence,
\[
    N \le 4\left\lfloor c\left(n + \log_2 \sqrt{5}\right) - 1 \right\rfloor \le 4\left\lceil c n \right\rceil,
\]
where $c = 1 / \log_2\left(\frac{\sqrt{5} + 1}{2}\right)$.

%% file: app_proof/sub_active.tex
\subsection{Step-dependent active windows}
\label{app:actwindow}

Fix an input $x$, let $(q_1, \ldots, q_{k-1})$ denote the quotient sequence produced by EEA, and define $b_i := \floor{\log_2 q_i}$ for each iteration $i$.
Under the four-phase schedule, each EEA iteration $i$ is expanded into $4(b_i + 1)$ steps. The total number of steps is $N = 4\sum_{i=1}^{k-1} (b_i + 1) \le N_{\max} := 4\ceil{cn}$, where $c = 1/\log_2\left(\frac{\sqrt{5}+1}{2}\right)$ (see Appendix \ref{app:step_num}).

For a global step index $T \in \{1, 2, \ldots, N_{\max}\}$, let $j = j(x, T)$ denote the index of the EEA iteration that contains step $T$, and let $u = u(x, T)$ denote the position of $T$ within that iteration. More precisely, the pair $(j(x,T), u(x,T))$ is uniquely determined by the conditions
\[
    \sum_{i=1}^{j-1} 4(b_i + 1) < T \le \sum_{i=1}^{j} 4(b_i + 1),
    \quad
    u = T - \sum_{i=1}^{j-1} 4(b_i + 1).
\]
The bit-length of the current quotient $\ell_q(x, T)$ depends deterministically on indices $(j, u)$:
\begin{equation}\label{eqn:phase_decompose}
\begin{split}
    \ell_q(x, T) & = \begin{cases}
        0 & u(x, T)\in (0, b_j + 1] \cup (3(b_j + 1), 4(b_j + 1)],\\
        u(x, T) - (b_j + 1) & u(x, T)\in (b_j + 1, 2(b_j + 1)],\\
        3(b_j + 1) - u & u(x, T)\in (2(b_j + 1), 3(b_j + 1)].
    \end{cases}
\end{split}
\end{equation}

We next state a lemma that characterizes the maximal possible value of $t_j$, conditioned on the total number of steps $N_j$ taken before reaching the $j$-th iteration of EEA. The proof follows exactly the same adjustment argument as in Appendix \ref{app:step_num}, and is therefore omitted.
\begin{lemma}\label{log2t_j>phi*N/4}
    Suppose that $N_j = \sum_{i=1}^{j-1} 4(b_i + 1)$ is fixed. Then, over all possible EEA iteration indices $j$ and all admissible quotient sequences $(q_1, \ldots, q_{j-1})$, we have $t_j \ge F_{N_j/4 + 1}$, and consequently,
    \[
        \floor{\log_2 t_j} + 1 \ge \log_2\left(\frac{\sqrt{5}+1}{2}\right)\cdot \frac{N_j}{4}.
    \]
    Here, $\{F_k\}_{k\ge 0}$ denotes the Fibonacci sequence $F_0 = 0, F_1 = 1$, and $F_{k+2} = F_{k+1} + F_k$ for all $k\ge 0$.
\end{lemma}

We then provide the detailed proofs of the bounds stated in Section \ref{subsec:actwindow} in the remainder of this subsection. Recall that $c = 1 / \log_2\left(\frac{\sqrt{5}+1}{2}\right)$.

\paragraph{Location-controlled addition/subtraction on $r$'s.}
First, observe that $t_j q_j < t_{j+1} < p$, which implies $\ell_t + b_j \le n + 1$. Since a location-controlled addition/subtraction on $r$'s is only activated in Phase 1 and 2, i.e. $0 < u\le 2(b_j + 1)$, we then analyze the claimed bound by considering two cases.
\begin{itemize}
    \item \textbf{Case 1.} Suppose that $u \le b_j + 1$. In this case, we have $\ell_q = 0$. By Lemma \ref{log2t_j>phi*N/4}, it follows that
    \[
        \ell_t \ge \log_2\left(\frac{\sqrt{5}+1}{2}\right)\cdot\frac{T - u}{4}
        \ge \log_2\left(\frac{\sqrt{5}+1}{2}\right)\cdot\frac{T - (n + 2 - \ell_t)}{4},
    \]
    hence $\ell_t + \ell_q = \ell_t \ge \dfrac{T - n - 2}{4c - 1}$.

    \item \textbf{Case 2.} Suppose that $b_j + 1 < u \le 2(b_j + 1)$. Then $\ell_q = u - (b_j + 1)$, and we obtain that
    \[
        \ell_t + \ell_q \ge \log_2\left(\frac{\sqrt{5}+1}{2}\right)\cdot\frac{T - u}{4} + (u - b_j - 1).
    \]
    The right-hand side is minimized when $u = b_j + 1$, in which case the expression reduces to that of Case 1.
\end{itemize}
Combining the two cases, we conclude that $\ell_t + \ell_q + 2 \ge \max\left\{\ceil{\dfrac{T - n - 2}{4c - 1}}, 1\right\} + 2$.

\paragraph{Location-controlled swaps.}
For the lower bound, a location-controlled swap is only activated in Phase 2 and 3, i.e. $b_j + 1 < u\le 3(b_j + 1)$, we then analyze the claimed lower bound by considering two cases.
\begin{itemize}
    \item \textbf{Case 1.} Suppose that $b_j + 1 < u \le 2(b_j + 1)$. In this case, we also have $\ell_t + \ell_q \ge \dfrac{T - n - 2}{4c - 1}$.

    \item \textbf{Case 2.} Suppose that $2(b_j + 1) < u \le 3(b_j + 1)$. In this case, we have $\ell_q = 3(b_j + 1) - u$. By Lemma \ref{log2t_j>phi*N/4}, it follows that
    \[
        \ell_t \ge \log_2\left(\frac{\sqrt{5}+1}{2}\right)\cdot\frac{T - u}{4}
        \ge \log_2\left(\frac{\sqrt{5}+1}{2}\right)\cdot\frac{T - 3(n + 2 - \ell_t)}{4},
    \]
    hence $\ell_t + \ell_q \ge \ell_t \ge \dfrac{T - 3(n + 2)}{4c - 3}$.
\end{itemize}
Combining the two cases, we conclude that $\ell_t + \ell_q + 1 \ge \max\left\{\ceil{\dfrac{T - 3(n + 2)}{4c - 3}}, 1\right\} + 1$.

For the upper bound, recall the recurrence relation of the EEA on the $t$-sequence, that is $t_{i+1} = t_{i-1} + q_i t_i < (q_i + 1)t_i \le 2^{b_i + 1} t_i$.
By iterating this inequality, we obtain that
\[
    t = t_j \le \prod_{i=1}^{j-1} 2^{b_i + 1} = 2^{(T - u)/4}, 
    \quad \text{and then} \quad 
    \ell_t \le (T - u) / 4 + 1.
\]
By Equation \eqref{eqn:phase_decompose}, we know that $\ell_q \le u/2$. Hence,
\[
    \ell_t + \ell_q + 1 \le \left\lfloor (T - u)/4 + u/2 + 2 \right\rfloor
    = \left\lfloor (T + u)/4 \right\rfloor + 2
    \le \left\lfloor T/2 \right\rfloor + 2.
\]
Finally, since the \texttt{Work1} register stores a value $r \ge 1$, we conclude that $\ell_t + \ell_q + 1 \le n + 2$.

\paragraph{Location-controlled addition/subtraction on $t$'s.}
As we proved before, We have that $\ell_t + 1 \le \left\lfloor (T - u)/4 \right\rfloor + 2\le \left\lceil T/4 \right\rceil + 1$. By the constraint that $t\le p < 2^n$, we also obtain that $\ell_t + 1\le n + 1$.

\paragraph{Length-update circuit.}
For the lower bound $k_4(T)$, the length-update circuit is activated only at the end of an EEA iteration, namely when $u = 4(b_j + 1)$. By Lemma \ref{log2t_j>phi*N/4}, we have
\[
    \ell_t \ge \log_2\left(\frac{\sqrt{5}+1}{2}\right)\cdot\frac{T - u}{4}
    \ge \log_2\left(\frac{\sqrt{5}+1}{2}\right)\cdot\frac{T - 4(n + 2 - \ell_t)}{4},
\]
which implies $\ell_t \ge \dfrac{T - 4(n + 2)}{4c - 4}$.

For the lower bound $k_5(T)$, Lemma \ref{log2t_j>phi*N/4} directly yields
\[
    \ell_t^* \ge \log_2\left(\frac{\sqrt{5}+1}{2}\right)\cdot \frac{T}{4} = \frac{T}{4c}.
\]

For the upper bound $K_5(T)$, recall the recurrence relation of the EEA on the $r$-sequence, $r_{i-1} = r_{i+1} + q_i r_i < (q_i + 1)r_i \le 2^{b_i + 1} r_i$.
By iterating this inequality, we obtain
\[
    p / r_{j+1} \le \prod_{i=1}^{j} 2^{b_i + 1} = 2^{T/4}.
\]
It follows that $n + 4 - \ell_{r'}^* < n + 4 - \log_2 r_{j+1} \le T/4 + n + 4 - \log_2 p \le T/4 + 5$, and hence $n + 4 - \ell_{r'}^* \le T/4 + 4$.

For the upper bound $K_4(T)$, we similarly have
$n + 3 - \ell_{r'}\le n + 3 - \ell_{r'}^* \le T/4 + 3$.

%% file: app_anal/app_anal.tex
\section{Optimization of Quantum Constant Adder}
\label{app:adder}

We adopt the linear-depth ripple-carry adder of Cuccaro et al. \cite{cuccaro2004new}, which computes carries in a
forward sweep and erases them in a backward sweep using two elementary reversible blocks: an in-place majority gate MAJ and an ``unmajority-and-add'' gate UMA. The MAJ and UMA blocks can both be implemented by two CNOT gates and one Toffoli gate.

Let $a = \sum_{i=0}^{n-1} a_i 2^i$ be an $n$-bit addend and $b = \sum_{i=0}^{n-1} b_i 2^i$ be an $n$-bit quantum integer. We denote the carry bits by $c_0, c_1, \ldots, c_n$ with $c_0 = 0$.
The classical carry recursion is
\[
    c_{i+1} = \mathrm{MAJ}(a_i,b_i,c_i), \qquad s_i = a_i \oplus b_i \oplus c_i,
\]
where $\mathrm{MAJ}(x,y,z)$ is the majority function. 
The Cuccaro construction computes $c_1, \ldots, c_n$ by chaining MAJ blocks, and then uncomputes carries while writing the sum bits by chaining UMA blocks. 

\paragraph{Gate count for the baseline.}
Each bit position contributes one MAJ block and one UMA block. With MAJ realized as $(2 \mathrm{CNOT} + 1\mathrm{Toffoli})$ and UMA (2-CNOT version) realized as $(2 \mathrm{CNOT} + 1 \mathrm{Toffoli})$, the resulting ripple-carry constant adder uses
\[
    \#\mathrm{Toffoli} = 2n, \qquad \#\mathrm{CNOT} = 4n + 1,
\]
where the additional $+1$ CNOT accounts for the final XOR into the most significant sum bit in the standard MAJ/UMA chaining pattern (matching the circuit structure in the ripple-carry diagram).

\paragraph{Specialize to the increment.}
We now specialize the constant adder to the increment operation $b \mapsto b+1$. In this case, the addend satisfies $a_0 = 1, a_i = 0$ for $i = 1, 2, \ldots, n - 1$.
This bit pattern simplifies both the carry recursion and the sum expressions, which in turn allows systematic cancellation of CNOT gates whose controls are zeros.
\begin{itemize}
\item For $i\ge 1$ we have $a_i = 0$, hence
\begin{equation}
    c_{i+1} = \mathrm{MAJ}(0, b_i, c_i) = b_i c_i.
\label{eq:carry-simplify}
\end{equation}
Operationally, any CNOT in the MAJ template whose control is $a_i$ becomes trivial and can be removed. What remains is exactly one Toffoli gate to compute $c_{i+1}$ from $(b_i, c_i)$ (written into
the designated carry location), with no CNOTs for these stages.

\item For the least significant bit, $a_0 = 1$ and $c_0 = 0$. The first carry is
\[
    c_1 = \mathrm{MAJ}(1, b_0, 0) = b_0,
\]
so the initialization of the carry chain can be implemented with only Clifford operations (a single CNOT that copies $b_0$ into the carry wire and two NOT gates).

\item For $i\ge 1$, since $a_i = 0$, the sum bit reduces to $s_i = b_i \oplus c_i$. In the backward sweep, UMA simultaneously (i) erases $c_{i+1}$ and (ii) writes $s_i$ into the target bit line. With $a_i = 0$, the CNOTs in the 2-CNOT UMA template that are controlled by $a_i$ again vanish. 
The stage can therefore be implemented using: (i) one CNOT to compute $b_i \leftarrow b_i \oplus c_i$ (thereby writing $s_i$), and (ii) one Toffoli to uncompute the carry contribution corresponding to Equation \eqref{eq:carry-simplify}.

\item The least significant bit can similarly be performed using a single CNOT that copies $b_0$ into the carry wire and one NOT gate.
\end{itemize}

Collecting the above simplifications yields the following count.

\paragraph{Gate count for the incrementer.}
Let $b$ be an $n$-bit quantum integer and consider the constant increment map $b \mapsto b+1$ implemented by specializing the Cuccaro ripple-carry adder with the 2-CNOT UMA construction. Then the circuit can be compiled to use
\[
    \#\mathrm{Toffoli} = 2n - 2, \qquad \#\mathrm{CNOT} = n + 2,
\]
up to Clifford-only single-qubit corrections (NOT gates).